\documentclass[aps,amsmath]{revtex4}
\usepackage{graphicx}  % needed for figures
\usepackage{dcolumn}   % needed for some tables
\usepackage{amssymb}   % for math
\usepackage{appendix}
\def\MET{{\mbox{$E\kern-0.57em\raise0.19ex\hbox{/}_{T}$}}}
\def\met{{\mbox{$E\kern-0.57em\raise0.19ex\hbox{/}_{T}$}}}
\def\DZ{D0 }
\def\DZero{D0 }
\def\Dzero{D0 }

\def\ifb{~fb$^{-1}$}
\def\pp{$p\bar{p}$}

\def\ttbar{$t\bar{t}$}

\def\lmet{$WH\rightarrow \ell\kern-0.45em\raise0.19ex\hbox{/} \nu b\bar{b}$}

\def\vww{$VH \rightarrow \ell^{\pm}\ell'^{\pm} + X$}

\def\hww{$H\rightarrow W^+ W^-$}
\def\hbb{$H\rightarrow b\bar{b}$}
\newcommand{\tm}[1]{\textrm{#1}}

\def\tevE{$\sqrt{s}=1.96$~TeV}

\begin{document}
%\linenumbers

\rightline{FERMILAB-CONF-11-044-E}
\rightline{CDF Note 10441}
\rightline{\DZ Note 6184}
\vskip0.5in

\title{Combined CDF and \DZ Upper Limits on Standard Model Higgs Boson Production with up to 8.2 fb$^{-1}$ of Data\\[2.5cm]}

\author{The TEVNPH Working Group\footnote{The Tevatron
New-Phenomena and Higgs Working Group can be contacted at
TEVNPHWG@fnal.gov. More information can be found at http://tevnphwg.fnal.gov/.}}

\affiliation{\vskip0.3cm for the CDF and \DZ Collaborations\\ \vskip0.2cm
\today}
\begin{abstract}
\vskip0.3in
We combine results from CDF and D0's direct searches for the standard model (SM)
Higgs boson ($H$) produced in \pp~collisions at the Fermilab Tevatron at $\sqrt{s}=1.96$~TeV.
The results presented here include those channels which are most sensitive to Higgs bosons with mass
between 130 and 200~GeV/$c^2$, namely searches targeted at Higgs boson
decays to $W^+W^-$, although
acceptance for decays into $\tau^+\tau^-$ and $\gamma \gamma $ is included.
Compared to the previous Tevatron Higgs search combination,
more data have been added and the analyses have been improved to gain sensitivity.
We use the MSTW08 parton distribution
functions and the latest $gg \rightarrow H$ theoretical cross section predictions
when testing for the presence of a SM Higgs boson.  With up to 7.1\ifb\ of data analyzed at CDF, and up
to 8.2\ifb\ at D0,
the 95\% C.L. upper limits on Higgs boson production is a factor of 0.54
times the SM cross section for a Higgs boson mass of 165~GeV/$c^2$.
We exclude at the 95\% C.L. the region $158<m_{H}<173$~GeV/$c^{2}$.
\\[2cm]
{\hspace*{5.5cm}\em Preliminary Results}
\end{abstract}

\maketitle

\newpage
%\vspace*{0.1cm}
%%%%%%%%%%%%%%%%%%%%%%%%%%%%%%%%%%%%%%%%%%%%%%%%%%%%%%%%%%%%%%%%%%%%%%%
%%%%%%%%%%%%%%%%%%%%%%%%%%%%%%%%%%%%%%%%%%%%%%%%%%%%%%%%%%%%%%%%%%%%%%%
\section{Introduction} %%%%%%%%%%%%%%

The search for a mechanism for electroweak symmetry breaking, and in
particular for a standard model (SM) Higgs boson, has been a major
goal of particle physics for many years, and is a central part of the
Fermilab Tevatron physics program, and also that of the Large Hadron Collider.
Recently, the ATLAS and CMS
collaborations at the Large Hadron Collider have released results on searches for the
Standard Model Higgs boson 
decaying to $W^+W^-$~\cite{atlasww,cmsww}, $ZZ$~\cite{atlaszz}, $\gamma\gamma$~\cite{atlasgammagamma}
Both the CDF and \Dzero collaborations
have performed new combinations~\cite{cdfHWW,DZHiggs} of multiple
direct searches for the SM Higgs boson.
The new searches cover a larger data sample, and incorporate
improved analysis techniques compared to previous analyses.
The sensitivities of these new combinations significantly exceed those
of previous combinations~\cite{prevhiggs,WWPRLhiggs}.  

In this note, we combine the most recent results of
searches for the SM Higgs boson produced in \pp~collisions at~\tevE.  The analyses combined
here seek signals of Higgs bosons produced in association with
vector bosons ($q\bar{q}\rightarrow W/ZH$), through gluon-gluon
fusion ($gg\rightarrow H$), and through vector boson fusion (VBF)
($q\bar{q}\rightarrow q^{\prime}\bar{q}^{\prime}H$) corresponding
to integrated luminosities up to 7.1\ifb~at CDF and up to
8.2\ifb~at D0.  
The searches included here target Higgs boson decays to $W^+W^-$, although
acceptance for decays into $\tau^+\tau^-$ and $\gamma \gamma $ are included in the D0 channels.

The searches are separated into
46 mutually exclusive final states (12 for CDF and 34 for D0;
see Tables~\ref{tab:cdfacc} and~\ref{tab:dzacc}) referred to
as ``analysis sub-channels'' in this note.  The selection procedures for each
analysis are detailed in Refs.~\cite{cdfHWW} through~\cite{dzHgg},
and are briefly described below.

%%%%%%%%%%%%%%%%%%%%%%%%%%%%%%%%%%%%%%%%%%%%%%%%%%%%%%%%%%%%%%%%%%%%%%%
\section{Channels Included in the Combination}
\label{sec:channels}

Event selections are similar for the corresponding CDF and D0 analyses.

\smallskip

For the \hww~analyses which seek events in which both $W$ bosons decay leptonically,
signal events are characterized by large \met~and
two oppositely-signed, isolated leptons.  The presence of neutrinos in the
final state prevents the accurate reconstruction of the candidate Higgs boson mass.
D0 selects events containing large \met\ and electrons and/or muons, dividing the data sample
into three final states: $e^+e^-$, $e^\pm \mu^\mp$, and $\mu^+\mu^-$.
The searches in the three final states each use 8.1 fb$^{-1}$ of data which is subdivided according to
the number of jets in the event: 0, 1, or 2 or more (``2+'') jets.
%Ref.~\cite{dzHWW1}.
Decays involving tau leptons are included in two orthogonal ways. For the first time, a dedicated
analysis ($\mu\tau_{had}$) using 7.3 fb$^{-1}$
of data studying the final state involving a muon and a hadronic tau decay, is included in the
Tevatron combination. Final states involving
other tau decays and mis-identified hadronic tau decays are included in the $e^+e^-$, $e^\pm \mu^\mp$,
and $\mu^+\mu^-$ final state analyses.
CDF separates the \hww\ events in five non-overlapping samples, split into
both ``high $s/b$'' and ``low $s/b$'' categories based on lepton types and
different categories based on the number of reconstructed jets: 0, 1, or 2+
jets.  The sample with two or more jets is not split into low $s/b$ and
high $s/b$ lepton categories due to low statistics.  A sixth CDF channel
is the low dilepton mass ($m_{\ell^+\ell^-}$) channel, which accepts events
with $m_{\ell^+\ell^-}<16$~GeV.  This channel increases the sensitivity of
the $H\rightarrow W^+W^-$ analyses at low $m_H$, adding 10\% additional
acceptance at $m_H=120$~GeV.
The division of events into jet categories allows the analysis discriminants
to separate three different categories of signals from the backgrounds more
effectively.  The signal production mechanisms considered are $gg\rightarrow
H\rightarrow W^+W^-$, $WH+ZH\rightarrow jjW^+W^-$, and vector-boson fusion.
The D0 $e^+e^-$, $e^\pm \mu^\mp$, and $\mu^+\mu^-$ final state channels use
boosted decision tree outputs as the final discriminants while
the $\mu\tau_{had}$ channel uses neural networks.
CDF uses neural-network outputs, including likelihoods constructed from
calculated matrix-element probabilities as additional inputs for the 0-jet bin.

D0 includes \vww\ analyses in which the associated vector boson and the $W$ boson from the Higgs
boson decay that has the same charge are required to decay leptonically,
thereby defining three like-sign dilepton final states ($e^\pm e^\pm$, $e^\pm
\mu^\pm$, and $\mu^{\pm}\mu^{\pm}$).  The combined output of two decision
trees, trained against the instrumental and diboson backgrounds respectively,
is used as the final discriminant.  CDF also includes a separate analysis
of events with same-sign leptons and large \met\ to incorporate additional
potential signal from associated production events in which the two leptons
(one from the associated vector boson and one from a $W$ boson produced in the
Higgs decay) have the same charge.  CDF for the first time also incorporates
three tri-lepton channels to include additional associated production
contributions where leptons result from the associated $W$ boson and the two
$W$ bosons produced in the Higgs decay or where an associated $Z$ boson decays
into a dilepton pair and a third lepton is produced in the decay of either
of the $W$ bosons resulting from the Higgs decay.  In the latter case, CDF
separates the sample into 1 jet and 2+ jet sub-channels to fully take
advantage of the Higgs mass constraint available in the 2+ jet case where
all of the decay products are reconstructed.

CDF also includes opposite-sign channels in which one of the
two lepton candidates is a hadronic tau.  Events are separated into $e$-$\tau$
and $\mu$-$\tau$ channels.  The final discriminants are obtained from boosted
decision trees which incorporate both hadronic tau identification and kinematic
event variables as inputs.

D0 includes channels in which one of the $W$ bosons in the $H \rightarrow W^+W^-$
process decays leptonically and the other decays hadronically.
Electron and muon final states
are studied separately, each with 5.4 fb$^{-1}$ of data.  Random forests
are used for the final discriminants.

D0 has updated its search for
Higgs boson production in which the Higgs decays into a
pair of photons to include 8.2 fb$^{-1}$ of data and to use a boosted decision tree as the final
discriminant.

The D0 analysis of the $\tau^+\tau^-+2$~jets  final state
has also been updated to include additional data (4.3 fb$^{-1}$) and both muonic and electronic tau decays.
%A neural network output is used as the discriminant variable for
%RunIIa (the first 1.0 fb$^{-1}$ of data), while a boosted decision tree output
%is used for later data (RunIIb).
The output of boosted decision trees is used as the final discriminant.

For both CDF and D0, events from QCD multijet (instrumental) backgrounds
are measured in independent data samples using different methods.
For CDF, backgrounds from SM processes with electroweak gauge bosons or
top quarks were generated using \textsc{pythia}~\cite{pythia}, \textsc{alpgen}~\cite{alpgen},
\textsc{mc@nlo}~\cite{MC@NLO}, and \textsc{herwig}~\cite{herwig} programs.
For D0, these backgrounds were generated using \textsc{pythia},
\textsc{alpgen}, and \textsc{comphep}~\cite{comphep}, with \textsc{pythia}
providing parton-showering and hadronization for all the generators.  These
background processes were normalized using either experimental data or
next-to-leading order calculations (including \textsc{mcfm}~\cite{mcfm} for
the $W+$ heavy flavor process).

Tables~\ref{tab:cdfacc} and~\ref{tab:dzacc} summarize, for CDF and D0 respectively,
the integrated luminosities, the Higgs boson mass ranges over which the searches are performed,
and references to further details for each analysis.

%%%%%%%%%%%%%%%%%%%%%%%%%%%%%%%%%%%%%%%%%%%%%%%%%%
%%%%  CDF Numbers
%%%%%%%%%%%%%%%%%%%%%%%%%%%%%%%%%%%%%%%%%%%%%%%%%%

\begin{table}[h]
\caption{\label{tab:cdfacc}  Luminosity, explored mass range and references
for the different processes
and final states ($\ell=e, \mu$) for the CDF analyses.
}
\begin{ruledtabular}
\begin{tabular}{lccc} \\
Channel & Luminosity (fb$^{-1}$) & $m_H$ range (GeV/$c^2$) & Reference \\ \hline
$H\rightarrow W^+ W^-$ \ \ \ 2$\times$(0,1 jets)+(2+ jets)+(low-$m_{\ell\ell}$)+($e$-$\tau_{had}$)+($\mu$-$\tau_{had}$) & 7.1  & 130-200 & \cite{cdfHWW} \\
$WH \rightarrow WW^+ W^-$ \ \ \ (same-sign leptons 1+ jets)+(tri-leptons)                                               & 7.1  & 130-200 & \cite{cdfHWW} \\
$ZH \rightarrow ZW^+ W^-$ \ \ \ (tri-leptons 1 jet)+(tri-leptons 2+ jets)                                               & 7.1  & 130-200 & \cite{cdfHWW} \\
\end{tabular}
\end{ruledtabular}
\end{table}

\vglue 0.5cm

%%%%%%%%%%%%%%%%%%%%%%%%%%%%%%%%%%%%%%%%%%%%%%%%%%
%%%%  d0 Numbers
%%%%%%%%%%%%%%%%%%%%%%%%%%%%%%%%%%%%%%%%%%%%%%%%%%
\begin{table}[h]
\caption{\label{tab:dzacc}  Luminosity, explored mass range and references
for the different processes
and final states ($\ell=e, \mu$) for the D0 analyses.
%Most channels have an associated `$\times 2$' corresponding to the RunIIa and RunIIb subdivision.
Most analyses are in addition analyzed separately for Run~IIa and IIb.
%In some cases, not every sub-channel uses the same dataset, and a range of integrated
%luminosities is given.
}
\begin{ruledtabular}
\begin{tabular}{lccc} \\
Channel & Luminosity (fb$^{-1}$) & $m_H$ range (GeV/$c^2$) & Reference \\ \hline
%$H\rightarrow W^+ W^- \rightarrow e^\pm\nu e^\mp\nu, \mu^\pm\nu \mu^\mp\nu$ \ \ \    & 5.4  & 130-200 & \cite{dzHWW1}\\
%$H\rightarrow W^+ W^- \rightarrow e^\pm\nu \mu^\mp\nu$ \ \ \ (0,1,2+ jet)     & 6.7  & 130-200 & \cite{dzHWW2}\\
$H\rightarrow W^+ W^- \rightarrow \ell^\pm\nu \ell^\mp\nu$ \ \ \ (0,1,2+ jet)     & 8.1  & 130-200 & \cite{dzHWW}\\
$H\rightarrow W^+ W^- \rightarrow \mu\nu \tau_{had}\nu$ \ \ \      & 7.3  & 130-200 & \cite{dzHWWtau}\\
$H\rightarrow W^+ W^- \rightarrow \ell\bar{\nu} jj$      & 5.4  & 130-200 & \cite{dzHWWjj}\\
$VH \rightarrow \ell^\pm \ell^\pm\ + X $ \ \ \  & 5.3  & 130-200 & \cite{dzWWW} \\
%$H\rightarrow ZZ\rightarrow \ell^\pm \ell^\mp jj$ \ \ \  & 4.2  & 130-200 & \cite{dzZZlljj} \\
%$VH\rightarrow \tau^+\tau^- b\bar{b}/q\bar{q} \tau^+\tau^-$ \ \ \      & 5.3  & 130-200 & \cite{dzVHt1,dzVHt2} \\
$H$+$X$$\rightarrow$$ \ell^\pm \tau^{\mp}_{had}jj$  \ \ \      & 4.3  & 130-200 & \cite{dzVHt2} \\
$H \rightarrow \gamma \gamma$                                 & 8.2  & 130-150 & \cite{dzHgg} \\
\end{tabular}
\end{ruledtabular}
\end{table}

\section{Signal Predictions and Uncertainties}
\label{sec:theory}

We normalize our Higgs boson signal predictions to the most recent highest-order calculations
available, for all production processes considered.  The largest production cross section, $\sigma(gg\rightarrow H)$,
is calculated at next-to-next-to-leading order (NNLO) in QCD with soft gluon resummation to 
next-to-next-to-leading-log (NNLL) accuracy,
and also includes
two-loop electroweak effects and handling of the running $b$ quark
mass~\cite{anastasiou,grazzinideflorian}.  The numerical values in Table~\ref{tab:higgsxsec} are
updates~\cite{grazziniprivate} of these predictions with $m_t$ set to 173.1~GeV/$c^2$~\cite{tevtop10}, and an exact treatment
of the massive top and bottom loop corrections up to next-to-leading-order (NLO) + next-to-leading-log (NLL) accuracy.
The factorization and renormalization scale
choice for this calculation is $\mu_F=\mu_R=m_H$.
These calculations are
refinements of earlier NNLO calculations of the $gg\rightarrow H$
production cross section~\cite{harlanderkilgore2002,anastasioumelnikov2002,ravindran2003}.
Electroweak corrections were computed in Refs.~\cite{actis2008,aglietti}.
Soft gluon resummation was introduced in the prediction of the
$gg\rightarrow H$ production cross section in Ref.~\cite{catani2003}.

The $gg\rightarrow H$ production cross section depends strongly on
the gluon parton density function, and the accompanying value of
$\alpha_s(q^2)$.  The cross sections used here are calculated
with the MSTW 2008 NNLO PDF set~\cite{mstw2008}, as recommended by the PDF4LHC working group~\cite{pdf4lhc}.
The reason to use this PDF parameterization is that it is the only NNLO prescription that results from
a fully global fit to all relevant data.  We follow the PDF4LHC working group's prescription
to evaluate the uncertainties on the $gg\rightarrow H$ production cross section
due to the PDFs.  This prescription is to evaluate the predictions of $\sigma(gg\rightarrow H)$ at NLO
using the global NLO PDF sets CTEQ6.6~\cite{cteq66}, MSTW08~\cite{mstw2008}, and NNPDF2.0~\cite{nnpdf}, and to
take the envelope of the predictions and their uncertainties due to PDF+$\alpha_s$ for the three sets as the uncertainty range
at NLO.  The ratio of the NLO uncertainty range to that of just MSTW08 is then used to scale the
NNLO MSTW08 PDF+$\alpha_s$ uncertainty, to estimate a larger uncertainty at NNLO.  This procedure roughly doubles the
PDF+$\alpha_s$ uncertainty from MSTW08 at NNLO alone.  Other PDF sets, notably ABKM09~\cite{abkm09} and
HERAPDF1.0~\cite{herapdf}, have been proposed~\cite{bd,bagliocrit} as alternate PDF sets for use in setting
the uncertainties on $\sigma(gg\rightarrow H)$.  These two PDF sets fit HERA data only in the case of
HERAPDF1.0, and HERA, fixed-target deep-inelastic scattering (DIS), fixed-target Drell-Yan (DY), and neutrino-nucleon di-muon data
in the case of ABKM09, and do not
include Tevatron jet data~\cite{cdfinclusivejet}\cite{d0inclusivejet}.
 We note that HERA, fixed-target DIS, fixed-target DY, and neutrino-nucleon di-muon data do not probe directly
the gluon PDF at the scales most relevant for our analyses, but are most sensitive to the quark distributions.
The high-$E_T$ Tevatron jet data contain gluon-initiated
processes at the relevant momentum transfer scales.  The PDF constraints from the HERA, fixed-target DIS,
fixed-target DY, and di-muon data
are extrapolated to make predictions of high-$x$ gluons at the relevant energy scales.  The reliability of these
extrapolations is checked for the HERAPDF1.0 PDF set 
by showing the NLO prediction~\cite{fastnlo} of the inclusive jet spectrum 
as a function
of jet $E_T$ and $\eta$, as compared with measurements by CDF and D0 as shown 
in Figures~\ref{fig:d0herapdf} and~\ref{fig:cdfherapdf},
respectively.
The measured inclusive jet cross sections are significantly under-predicted using this PDF set.  The corresponding comparisons using the same
NLO jet cross section calculation with the MSTW08 PDF set are shown in Figures~49 and~50 in Ref.~\cite{mstw2008}.  The
agreement with the jet data is much better with the MSTW2008 PDF set, which is expected, since those data are included
in the MSTW08 PDF fit.  
Consequently the high-$x$ gluon component is larger in the MSTW2008 PDF set than in the
HERAPDF1.0 PDF set. This forms an important experimental foundation for the prediction of
the gluon-fusion Higgs boson production cross section.
Recently, it has been pointed out~\cite{abm2011} that the gluon PDF in the ABKM09 fit, and thus the values
of $\sigma(gg\rightarrow H)$, depend on the handling of the data input from the New Muon Collaboration (NMC), 
specifically whether the differential
spectrum as measured by NMC is included in the fit or whether the structure function $F_2$ reported by NMC is
used in the fit.  It is verified by the NNPDF Collaboration~\cite{nnpdfnmc} that the NNPDF2.1 global PDF fit
is negligibly impacted by changing the handling of the input NMC data, indicating that the global fit,
which includes Tevatron jet data, is more solid and less sensitive to the precise handling of NMC data.  It should be noted
however that this study is performed only at NLO.

We also include uncertainties on $\sigma(gg\rightarrow H)$  due to uncalculated higher order processes by following the
standard procedure of varying the factorization renormalization scales up and down together by a factor $\kappa=2$.
It had been proposed to vary $\mu_R$ and $\mu_F$ independently~\cite{bd}, but the maximum impact on $\sigma(gg\rightarrow H)$
is found when varying the two scales together.
At a central scale choice of $\mu_R=\mu_F=m_H/2$, the fixed-order calculations at NLO and NNLO converge rapidly and
the scale uncertainties generously cover the differences.  Furthermore, the NNLO calculation using a scale choice
of $\mu_R=\mu_F=m_H/2$ is very close to the NNLL+NNLO
calculation that we use.  The authors of the NNLL+NNLO calculation
recommend that we use the scale choice $\mu_R=\mu_F=m_H$ for the central value~\cite{grazzinideflorian}.
The uncertainty on $\sigma(gg\rightarrow H)$ is evaluated by varying $\mu_R=\mu_F$ from a lower value of $m_H/2$ to
an upper value of $2m_H$, the customary variation, and by evaluating the NNLO+NNLL cross sections at these scales.  
A larger scale variation has been proposed, varying the scale by a factor $\kappa=3$~\cite{bd}.  The original justification
for this proposal~\cite{bd} was for the LO calculation to cover the NNLO central value.  The
LO calculation however does not have enough scale-dependent terms for its scale uncertainty to be fairly considered -- it is missing
significant QCD radiation and other terms that increase the scale dependence at NLO compared with that at LO.  
The second justification~\cite{bagliocrit}
is simply that the higher-order QCD corrections are larger than those of other processes, such as Drell-Yan production.  
The corrections are large due to the
fact that there are gluons in the initial state which radiate more gluons more frequently than the quarks that 
initiate other processes.  The scale dependence of the calculations of $\sigma(gg\rightarrow H)$ which include
this radiation and other terms is correspondingly
larger, and the customary range of scale variation provides uncertainties which are commensurate with the values of the corrections.
An independent calculation by Ahrens, Becher, Neubert, and Yang~\cite{Ahrens} using a renormalization-group improved
resummation of radiative corrections at NNLL accuracy gives an alternative
calculation of $\sigma(gg\rightarrow H)$ that converges more rapidly and requires even smaller scale uncertainties.
The central values of this calculation of $\sigma(gg\rightarrow H)$ are in excellent agreement with those used in our analysis
and which are given in Table~\ref{tab:higgsxsec}.

Because the $H\rightarrow W^+W^-$ analyses separate the data into categories based on the numbers of observed jets,
we assess factorization and renormalization scale and PDF$+\alpha_s$ variations separately for each
jet category as evaluated in Ref.~\cite{anastasiouwebber}.  
This calculation is at NNLO for $H$+0~jets, at NLO for $H$+1~jet, and at LO for $H$+2 or more jets.  A newer,
more precise calculation~\cite{campbell2j}
of the $H$+2 or more jets cross section at NLO is used to evaluate the uncertainties in this category.
These scale uncertainties are used instead of the inclusive NNLL+NNLO scale uncertainty because we require
them in each jet category, and the uncertainties we use are significantly larger than the inclusive scale uncertainty.
The scale choice affects the $p_T$ spectrum of the Higgs boson when produced in gluon-gluon fusion, and this
effect changes the acceptance of the selection requirements and also the shapes of the distributions of the final
discriminants.  The effect of the acceptance change is included in the calculations of Ref.~\cite{anastasiouwebber}
and Ref.~\cite{campbell2j}, as the experimental requirements are simulated in these calculations.  
The effects on the
final discriminant shapes are obtained by reweighting the $p_T$ spectrum of the Higgs boson production in our
Monte Carlo simulation to higher-order calculations.  The Monte Carlo signal simulation used by CDF and D0
is provided by the LO generator {\sc pythia}~\cite{pythia} which includes a parton shower and fragmentation and
hadronization models.  We reweight the Higgs boson $p_T$ spectra in our {\sc pythia} Monte Carlo samples
to that predicted by {\sc hqt}~\cite{hqt} when making predictions of differential distributions of $gg\rightarrow H$ signal events.
To evaluate the impact of the scale uncertainty on our differential spectra, we use the {\sc resbos}~\cite{resbos} generator, and
apply the scale-dependent differences in the Higgs boson $p_T$ spectrum to the {\sc hqt} prediction, and propagate these to
our final discriminants as a systematic uncertainty on the shape, which is included in the calculation of the limits.

  We treat the scale uncertainties
as 100\% correlated between jet categories and between CDF and D0, and
also treat the PDF+$\alpha_s$ uncertainties in the cross section as correlated
between jet categories and between CDF and D0.  We treat however the PDF$+\alpha_s$ uncertainty
as uncorrelated with the scale uncertainty.  The main reason is that the PDF uncertainty arises
from experimental uncertainties and PDF parameterization choices, while the scale uncertainty
arises from neglected higher-order terms in the perturbative cross section calculations.
The PDF predictions do depend on the scale choice, however.
The dependence of the prediction of $\sigma(gg\rightarrow H)$ due to the scale
via the PDF dependence is included as part of the scale uncertainty and not as part of the PDF
uncertainty, to ensure that all scale dependence is considered correlated.
Furthermore, we have verified~\cite{anastasiouprivate} that the fractional change in the prediction of
$\sigma(gg\rightarrow H)$ due to the PDF variation depends negligibly on the value of the
scale choice, justifying the treatment of PDF and scale uncertainties as uncorrelated.  As described
in Section~\ref{sec:combination}, systematic uncertainties arising from uncorrelated sources should not
be added linearly together, but instead should be considered to fluctuate independently of one another.
We have investigated the use of a Gaussian prior for the scale uncertainty using the $\kappa=2$ variation
as the $\pm 1\sigma$ variation, and compared it with the use of a flat prior between the low edge and the
upper edge, and have found a negligible impact on our sensitivity and our observed limits.  The main reason
for this is that while the flat prior has a higher density near the edges, the Gaussian prior has longer tails, and
both effects essentially compensate.
The use of a flat prior with edges for the scale uncertainty does not affect the choice of whether to add this
uncertainty linearly or not with the PDF uncertainty; that choice is determined by the correlations of the
two sources of uncertainty.

Another source of uncertainty in the prediction of $\sigma(gg\rightarrow H)$ is the extrapolation of the
QCD corrections computed for the heavy top quark loops to the light-quark loops included as part of the electroweak
corrections.  Uncertainties at the level of 1-2\% are already included in the cross section values we
use~\cite{grazzinideflorian,anastasiou}.  In Ref.~\cite{anastasiou}, it is argued that the factorization of QCD
corrections is known to work well for Higgs boson masses many times in excess of the masses of the loop particles.
A 4\% change in the predicted cross section is seen when all QCD corrections are removed from the diagrams
containing light-flavored quark loops, which is too conservative.  For the $b$ quark loop, which is computed
separately in Ref.~\cite{anastasiou}, the QCD corrections are much smaller than for the top loop, further giving
confidence that it does not introduce large uncertainties.

\begin{figure}
\begin{center}
\includegraphics[width=\textwidth]{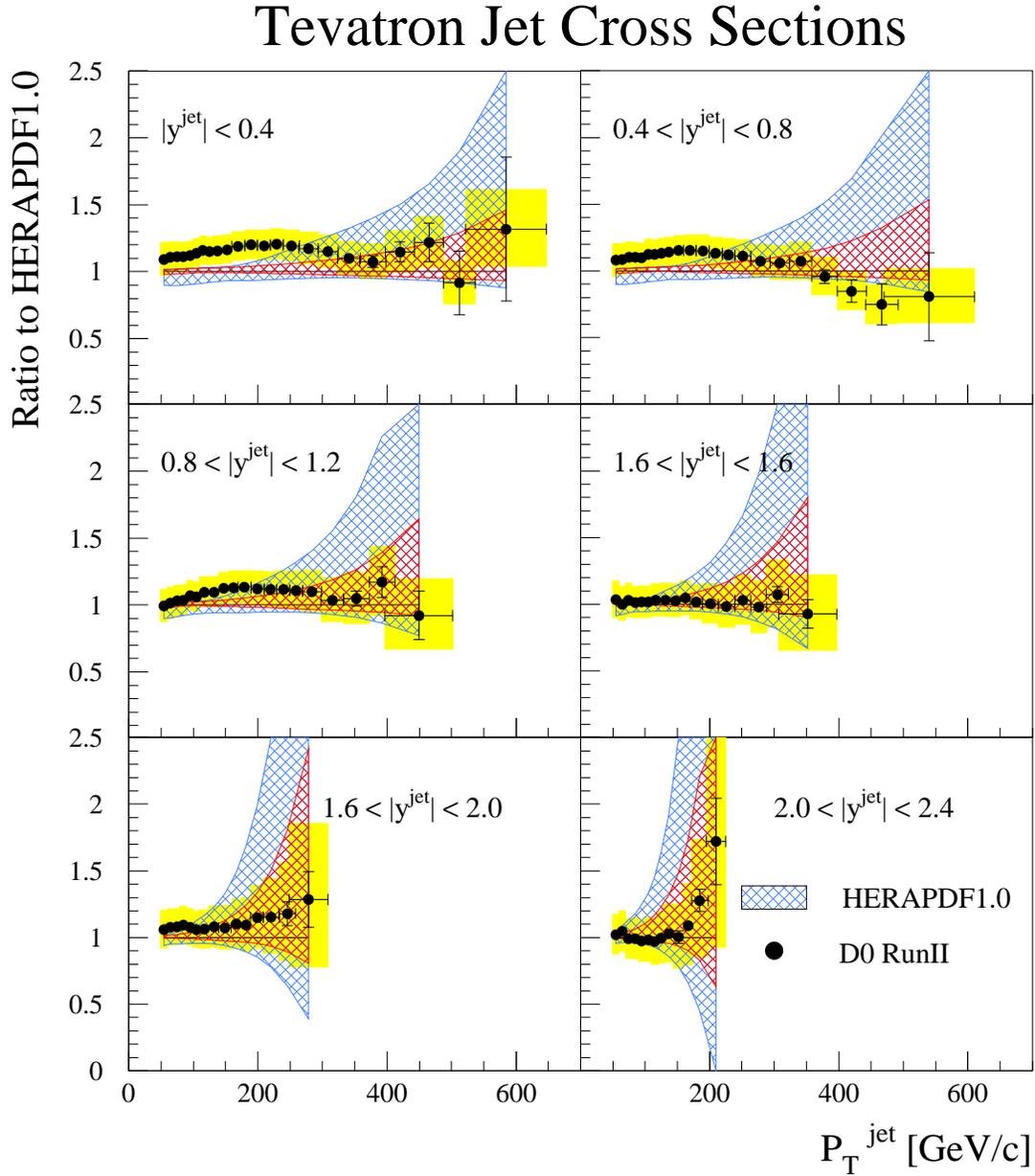}
\caption{\label{fig:d0herapdf}
Inclusive jet cross section measurements by D0 as a
function of jet $p_T$ in six bins of jet rapidity $y$, divided by the theoretical prediction.
The data systematic uncertainties are displayed
by the full shaded band. The theoretical prediction is an
NLO perturbative QCD calculation~\protect\cite{fastnlo}, with renormalization and
factorization scales set to jet $p_T$ using the HERAPDF1.0 PDFs~\protect{\cite{herapdf}\cite{herapdffigs}}
and including nonperturbative corrections.
The PDF uncertainties are shown as hatched regions -- the blue hatching shows the total PDF uncertainty,
and the red hatching shows the portion of the total due to experimental uncertainties.
This figure is obtained using the same
data and methods described in Ref.~\protect{\cite{d0inclusivejet}}.}
\end{center}
\end{figure}

\begin{figure}
\begin{center}
\includegraphics[width=\textwidth]{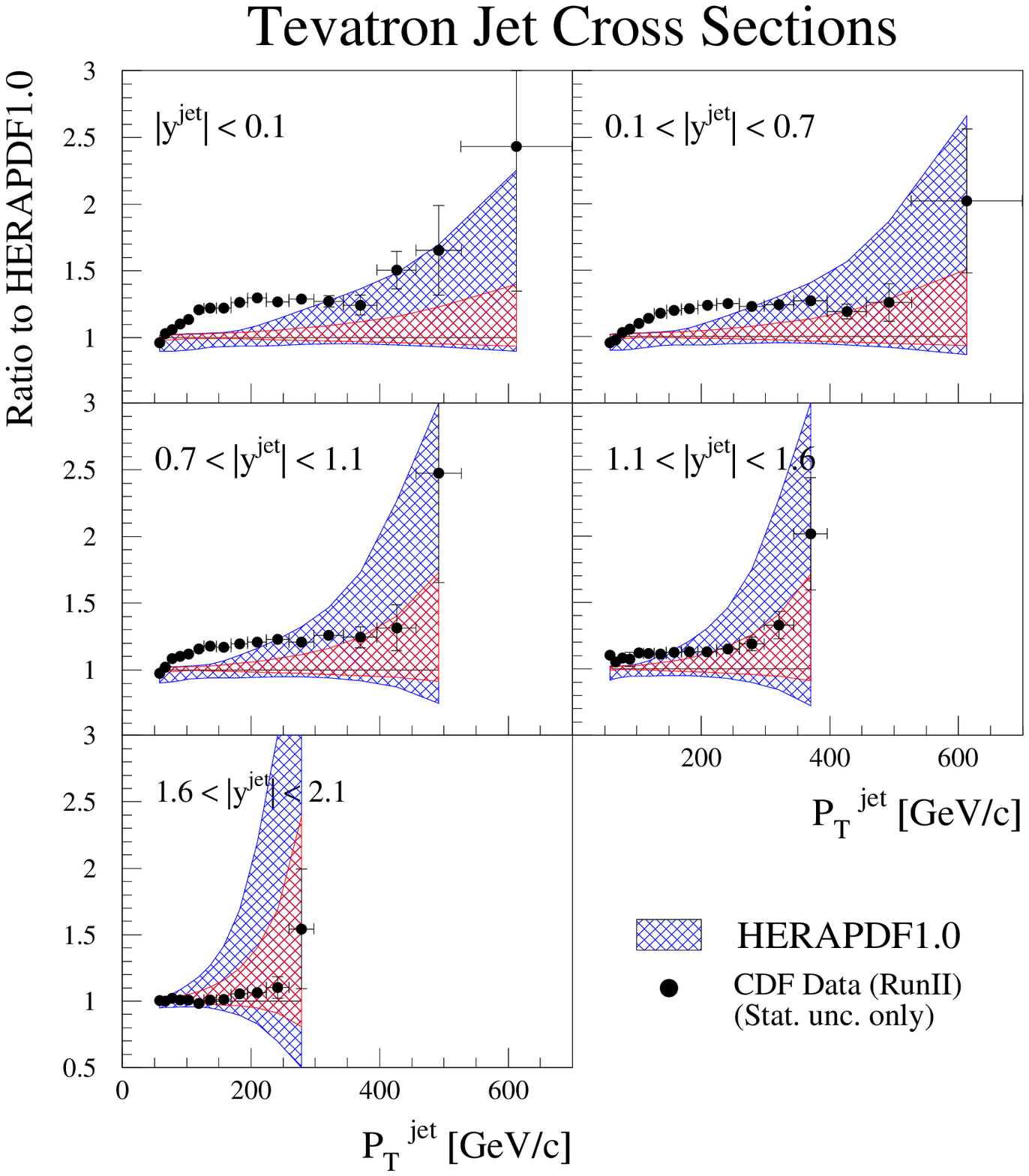}
\caption{\label{fig:cdfherapdf}
Inclusive jet cross section measurements by CDF as a 
function of jet $p_T$ in six bins of jet rapidity $y$, divided by the theoretical prediction.
The theoretical prediction is an
NLO perturbative QCD calculation~\protect\cite{fastnlo}, with renormalization and
factorization scales set to jet $p_T$ using the HERAPDF1.0 PDFs~\protect{\cite{herapdf}\cite{herapdffigs}}
and including nonperturbative corrections.
The PDF uncertainties are shown as hatched regions -- the blue hatching shows the total PDF uncertainty,
and the red hatching shows the portion of the total due to experimental uncertainties.  The
CDF data are published in Ref.~\protect{\cite{cdfinclusivejet}}.  Experimental uncertainties on the CDF
data are comparable to those of the D0 data, shown in Fig.~\protect{\ref{fig:d0herapdf}}.
}
\end{center}
\end{figure}

We include all significant Higgs boson production modes in our searches.
Besides gluon-gluon fusion through virtual quark loops
(GGF), we include Higgs boson production in association with a $W$
or $Z$ vector boson, and vector boson  fusion (VBF).  We use the $WH$ and $ZH$
production cross sections computed
at NNLO in Ref.~\cite{djouadibaglio}.  This calculation starts with the NLO calculation of
{\sc v2hv}~\cite{v2hv} and includes NNLO QCD contributions~\cite{vhnnloqcd}, as well
as one-loop electroweak corrections~\cite{vhewcorr}. 
We use the VBF cross section computed at NNLO in QCD in Ref.~\cite{vbfnnlo}.  Electroweak corrections
to the VBF production cross section are computed with the {\sc hawk} program~\cite{hawk}, and
are small and negative (-2\% to -3\%) for the Higgs boson mass range considered here.  They are not
included in this update but will be incorporated in future results.

In order to predict the distributions of the kinematics of Higgs boson signal events, CDF
and D0 use the \textsc{pythia}~\cite{pythia} Monte Carlo program,
with \textsc{CTEQ5L} and \textsc{CTEQ6L}~\cite{cteq} leading-order
(LO) parton distribution functions.  The Higgs boson decay branching
ratio predictions are calculated with \textsc{hdecay}~\cite{hdecay}, and are
also listed in Table~\ref{tab:higgsxsec}.  We use \textsc{hdecay} Version 3.53.
While the $HWW$ coupling is well predicted, $Br(H\rightarrow W^+W^-)$ depends on the
partial widths of all other Higgs boson decays. The partial width $\Gamma(H\rightarrow b{\bar{b}})$ is sensitive
to $m_b$ and $\alpha_s$, $\Gamma(H\rightarrow c{\bar{c}})$ is sensitive to $m_c$ and $\alpha_s$, and $\Gamma(H\rightarrow gg)$
is sensitive to $\alpha_s$.
The impacts of these uncertainties on $Br(H\rightarrow W^+W^-)$ depend on $m_H$ due to the fact that
$Br(H\rightarrow b{\bar{b}})$, $Br(H\rightarrow c{\bar{c}})$, $Br(H\rightarrow gg)$ become very small for Higgs boson
masses above 160~GeV$/c^2$, while they have a larger impact for lower $m_H$.  We use the uncertainties
on the branching fraction $Br(H\rightarrow W^+W^-)$ from Ref.~\cite{bagliodjouadilittlelhc}.
At $m_H=130$~GeV$/c^2$, for example, the $m_b$ variation gives a $^{-4.89}_{+1.70}\%$ relative variation
in $Br(H\rightarrow W^+W^-)$, $\alpha_s$ gives a $^{-1.02}_{+1.09}\%$ variation, and
$m_c$ gives a  $^{-0.45}_{+0.51}\%$ variation.  At $m_H=165$~GeV$/c^2$, all three of these
uncertainties are below 0.1\% and remain small for higher $m_H$.

% updated Feb. 22, 2010 -- ggh, wh, zh, vbf have new values. b.r.'s have not changed as of Jan. 2011
\begin{table}
\begin{center}
\caption{
The production cross sections and decay branching fractions
for the SM Higgs boson assumed for the combination.}
\vspace{0.2cm}
\label{tab:higgsxsec}
\begin{ruledtabular}
\begin{tabular}{ccccccccc}
$m_H$  & $\sigma_{gg\rightarrow H}$  & $\sigma_{WH}$  & $\sigma_{ZH}$  & $\sigma_{VBF}$  & $B(H\rightarrow \tau^+{\tau^-})$ & $B(H\rightarrow W^+W^-)$ & $B(H\rightarrow ZZ)$           & $B(H\rightarrow \gamma\gamma)$ \\ 
(GeV/$c^2$) & (fb)                   & (fb)           & (fb)           & (fb)            & (\%)                             & (\%)                     & (\%)                           & (\%) \\  \hline
   130 &    842.9   &   112.00    &    68.5      &     62.1 &    5.305 &  29.43 &  3.858 &  0.2182     \\
   135 &    750.8   &    97.20    &    60.0      &     57.5 &    4.400 &  39.10 &  5.319 &  0.2077     \\
   140 &    670.6   &    84.60    &    52.7      &     53.2 &    3.472 &  49.16 &  6.715 &  0.1897     \\
   145 &    600.6   &    73.70    &    46.3      &     49.4 &    2.585 &  59.15 &  7.771 &  0.1653     \\
   150 &    539.1   &    64.40    &    40.8      &     45.8 &    1.778 &  68.91 &  8.143 &  0.1357     \\
   155 &    484.0   &    56.20    &    35.9      &     42.4 &    1.057 &  78.92 &  7.297 &  0.09997    \\
   160 &    432.3   &    48.50    &    31.4      &     39.4 &    0.403 &  90.48 &  4.185 &  0.05365    \\
   165 &    383.7   &    43.60    &    28.4      &     36.6 &    0.140 &  95.91 &  2.216 &  0.02330    \\
   170 &    344.0   &    38.50    &    25.3      &     34.0 &    0.093 &  96.39 &  2.351 &  0.01598    \\
   175 &    309.7   &    34.00    &    22.5      &     31.6 &    0.073 &  95.81 &  3.204 &  0.01236    \\
   180 &    279.2   &    30.10    &    20.0      &     29.4 &    0.059 &  93.25 &  5.937 &  0.01024    \\
   185 &    252.1   &    26.90    &    17.9      &     27.3 &    0.046 &  84.50 &  14.86 &  0.008128   \\
   190 &    228.0   &    24.00    &    16.1      &     25.4 &    0.038 &  78.70 &  20.77 &  0.006774   \\
   195 &    207.2   &    21.40    &    14.4      &     23.7 &    0.033 &  75.88 &  23.66 &  0.005919   \\
   200 &    189.1   &    19.10    &    13.0      &     22.0 &    0.029 &  74.26 &  25.33 &  0.005285   \\
\end{tabular}	
\end{ruledtabular}	
\end{center}	
\end{table}

%%%%%%%%%%%%%%%%%%%%%%%%%%%%%%%%%%%%%%%%%%%%%%%%%%%%%%%%%%%%%%%%%%%%%%%
%%%%%%%%%%%%%%%%%%%%%%%%%%%%%%%%%%%%%%%%%%%%%%%%%%%%%%%%%%%%%%%%%%%%%%%
\section{Distributions of Candidates} %%%%%%%%%%%%%%

All analyses provide binned histograms of the final discriminant variables
for the signal and background predictions, itemized separately for each
source, and the observed data.
The number of channels combined is large, and the number of bins
in each channel is large.  Therefore, the task of assembling
histograms and checking whether the expected and observed limits are
consistent with the input predictions and observed data is difficult.
We thus provide histograms that aggregate all channels' signal,
background, and data together.  In order to preserve most of the
sensitivity gain that is achieved by binning the data
instead of collecting them all together and counting, we aggregate the
data and predictions in narrow bins of signal-to-background ratio,
$s/b$.  Data with similar $s/b$ may be added together with no loss in
sensitivity, assuming similar systematic errors on the predictions.
The aggregate histograms do not show the effects of systematic
uncertainties, but instead compare the data with the central
predictions supplied by each analysis.

The range of $s/b$ is quite large in each analysis, and so
$\log_{10}(s/b)$ is chosen as the plotting variable.  Plots of the
distributions of $\log_{10}(s/b)$ are shown for $m_H=150$ and 165~GeV/$c^2$ in Figure~\ref{fig:lnsb}.  These
distributions can be integrated from the high-$s/b$ side downwards,
showing the sums of signal, background, and data for the most pure
portions of the selection of all channels added together.  These
integrals can be seen in Figure~\ref{fig:integ}.  The most significant
candidates are found in the bins with the highest $s/b$; an excess
in these bins relative to the background prediction drives the Higgs
boson cross section limit upwards, while a deficit drives it downwards.
The lower-$s/b$ bins show that the modeling of the rates and kinematic
distributions of the backgrounds is very good.

We also show the distributions of the data after subtracting the
expected background, and compare that with the expected signal yield
for a standard model Higgs boson, after collecting all bins in all
channels sorted by $s/b$.  These background-subtracted distributions
are shown in Figure~\ref{fig:bgsub}.  These graphs also show the
remaining uncertainty on the background prediction after fitting the
background model to the data within the systematic uncertainties on
the rates and shapes in each contributing channel's templates.

 \begin{figure}[t]
 \begin{centering}
 \includegraphics[width=0.4\textwidth]{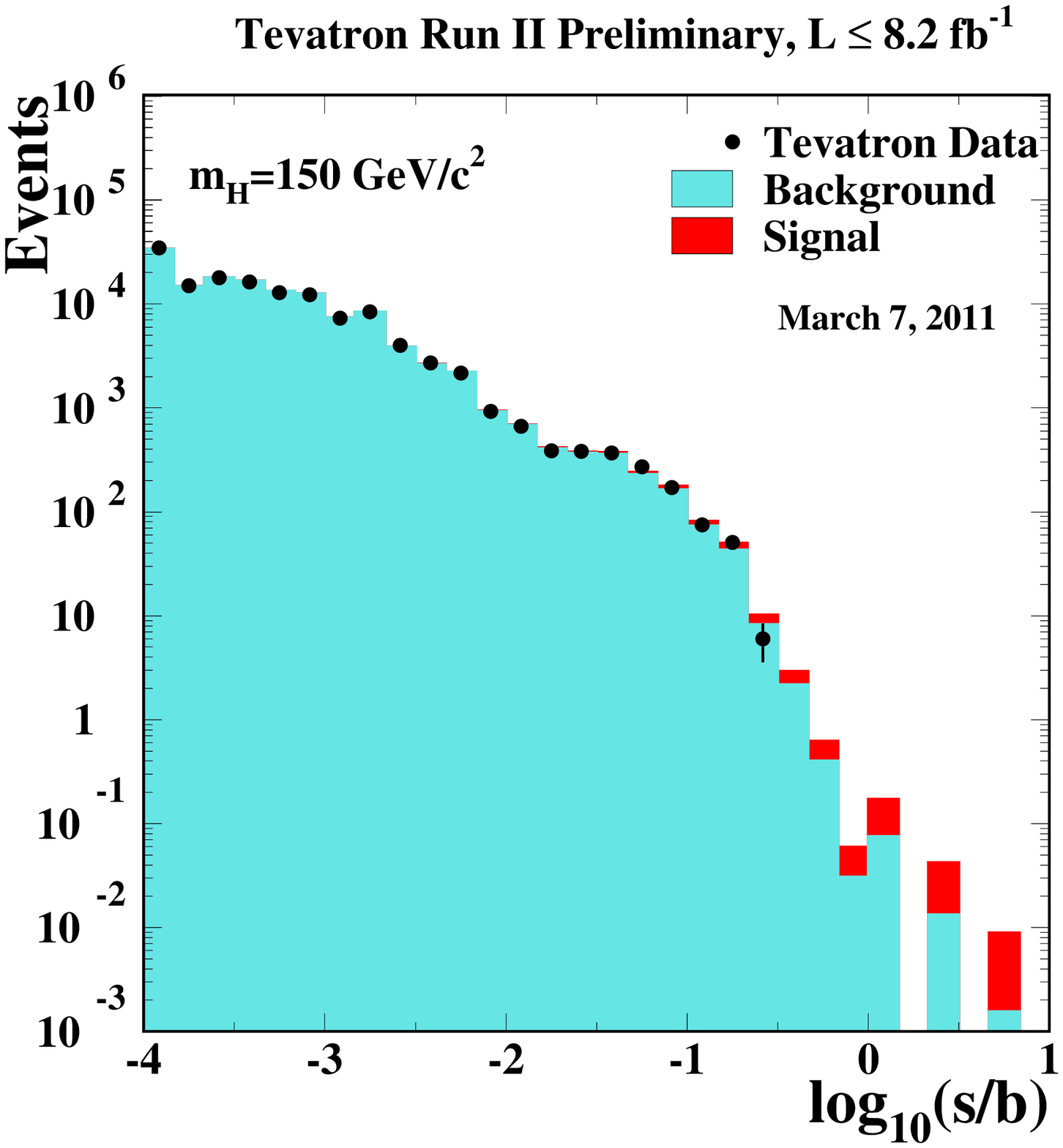}\includegraphics[width=0.4\textwidth]{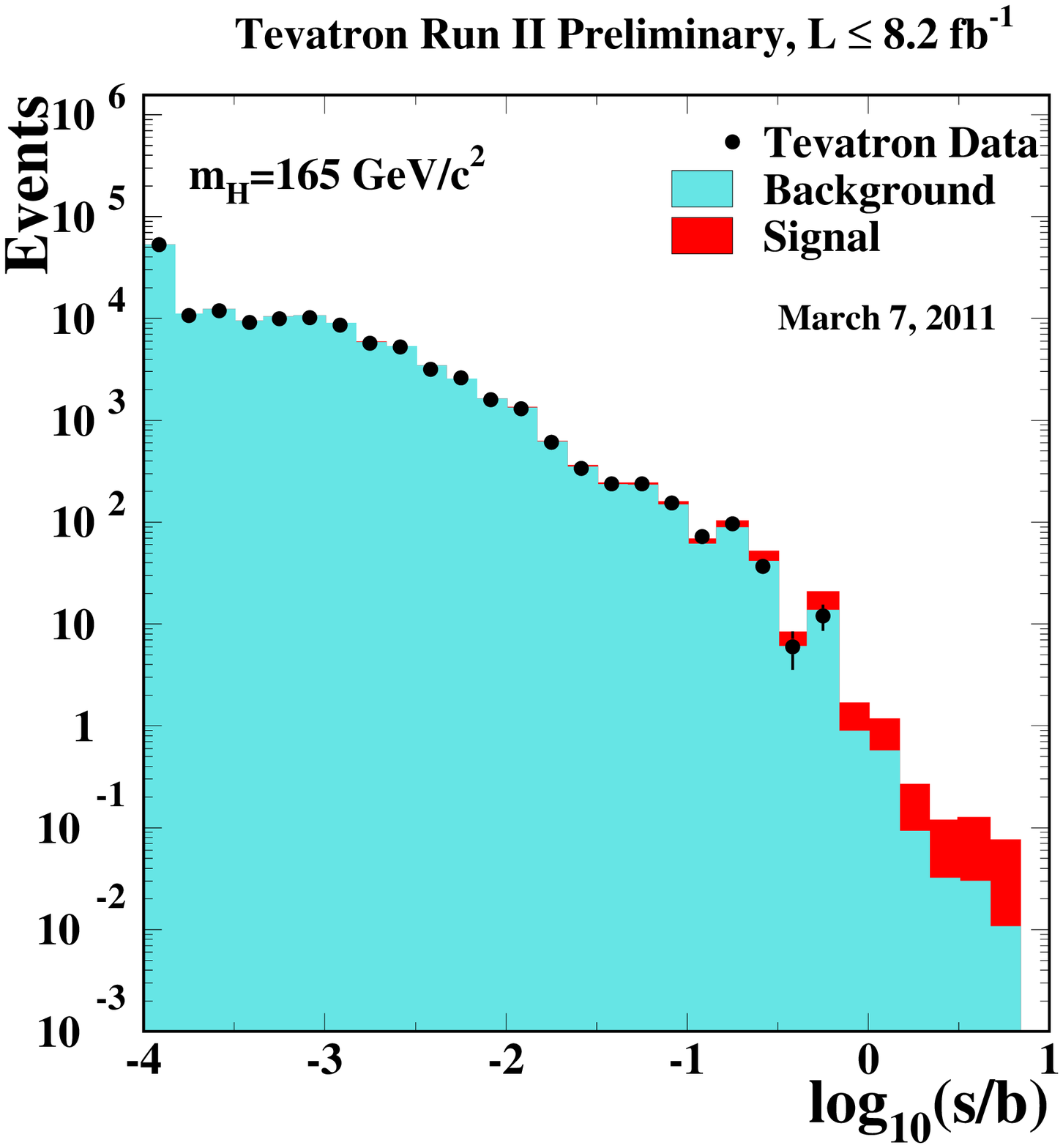}
 \caption{
 \label{fig:lnsb} Distributions of $\log_{10}(s/b)$, for the data from all contributing channels from
CDF and D0, for Higgs boson masses of 150 and 165~GeV/$c^2$.  The
data are shown with points, and the expected signal is shown stacked on top of
the backgrounds.  Underflows and overflows are collected into the
bottom and top bins. }
 \end{centering}
 \end{figure}

 \begin{figure}[t]
 \begin{centering}
 \includegraphics[width=0.4\textwidth]{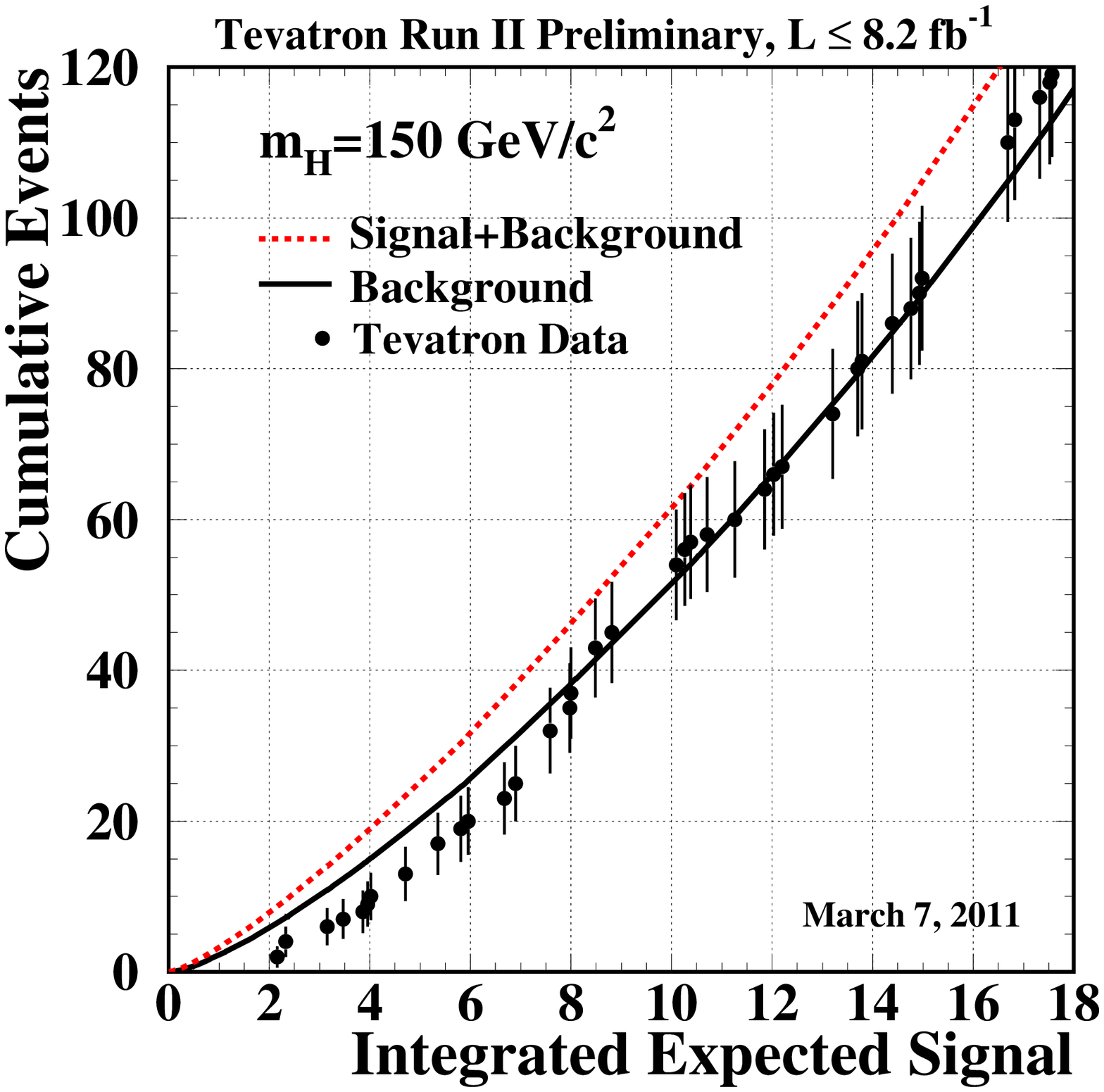}\includegraphics[width=0.4\textwidth]{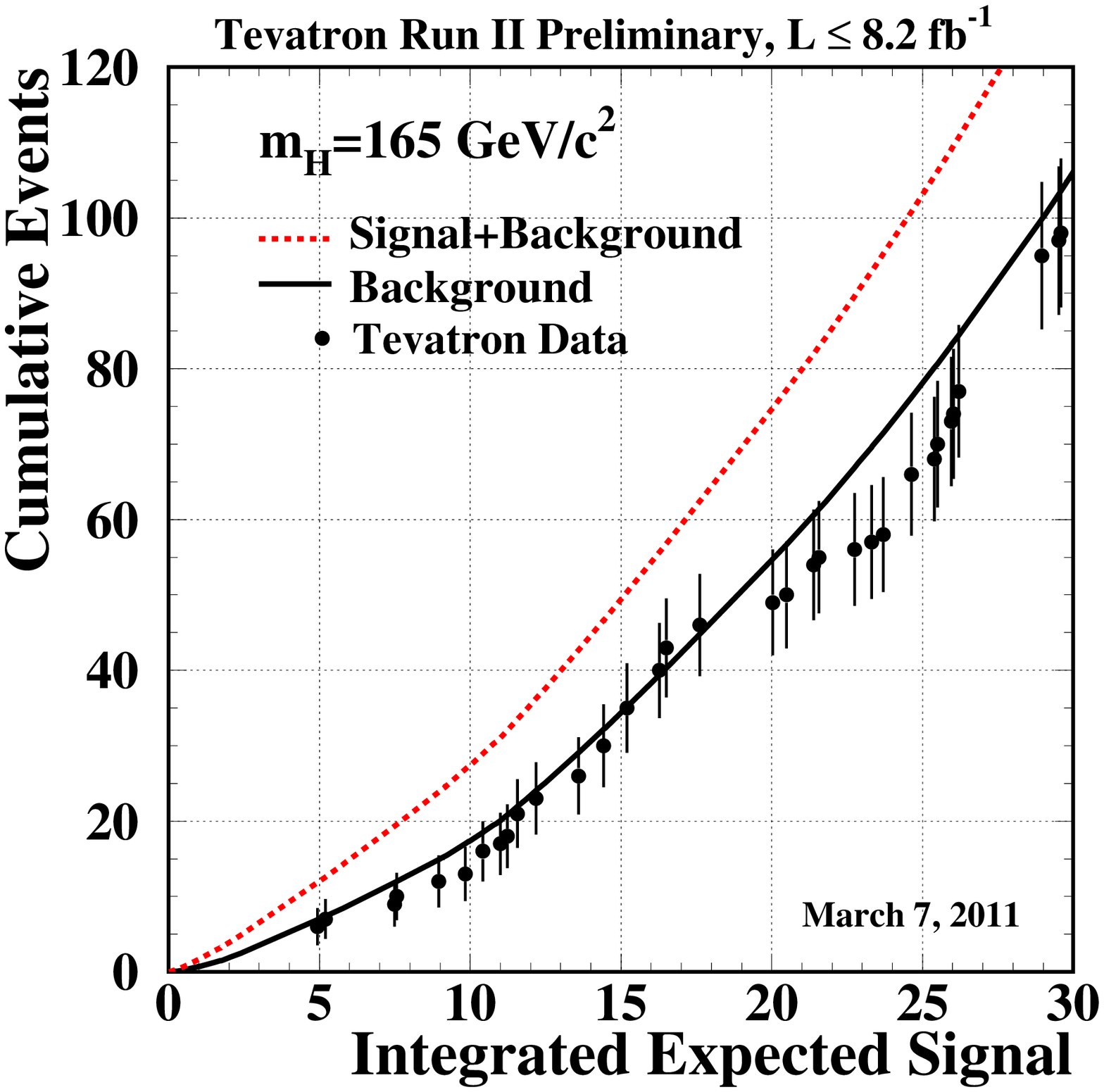}
 \caption{ 
 \label{fig:integ} Integrated distributions of $s/b$, starting at the high $s/b$ side, for Higgs boson
masses of 150 and 165~GeV/$c^2$.  The total signal-plus-background
and background-only integrals are shown separately, along with the
data sums.  Data are only shown for bins that have
data events in them.}
 \end{centering}
 \end{figure}

 \begin{figure}[t]
 \begin{centering}
 \includegraphics[width=0.45\textwidth]{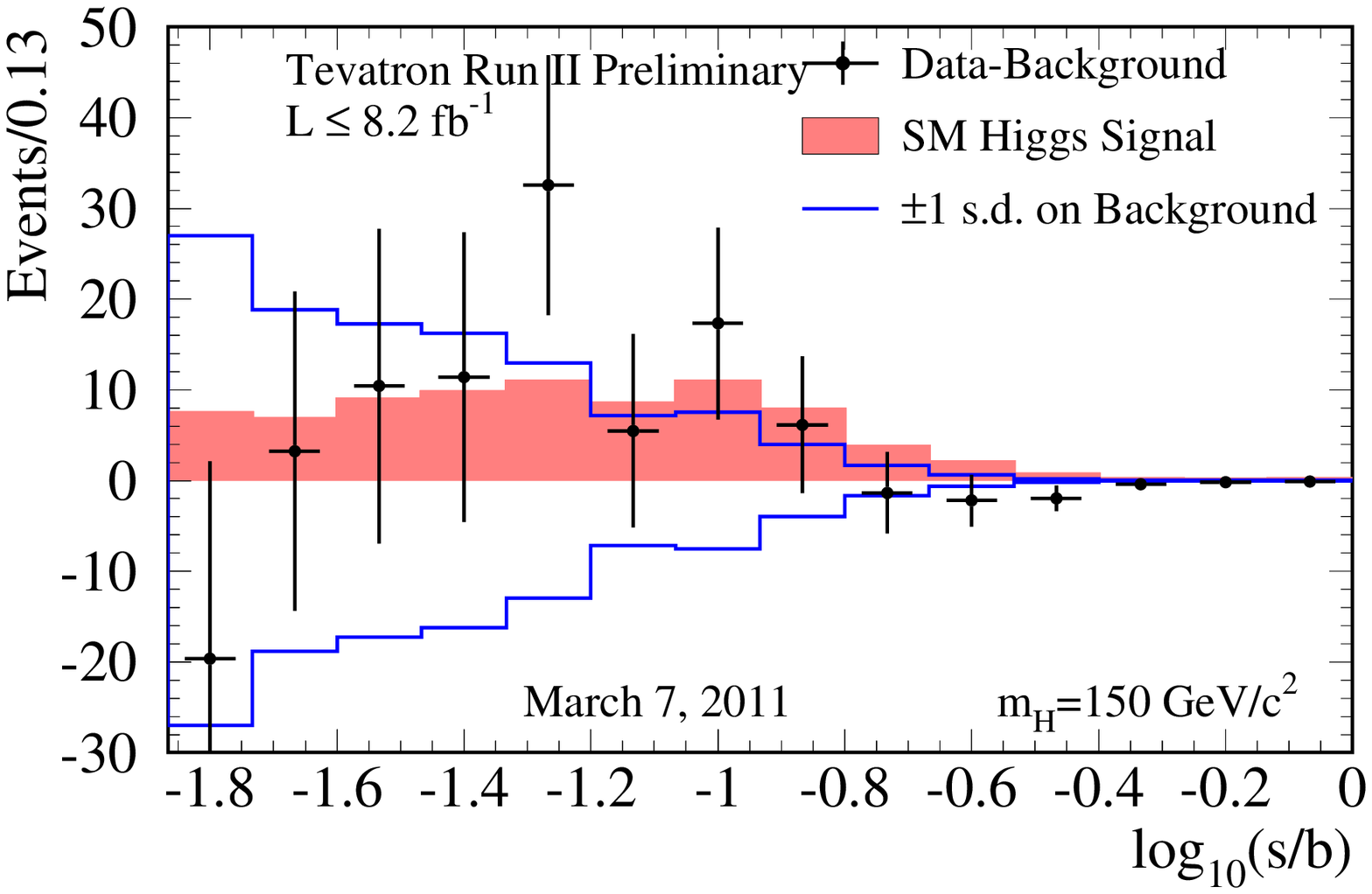}\includegraphics[width=0.45\textwidth]{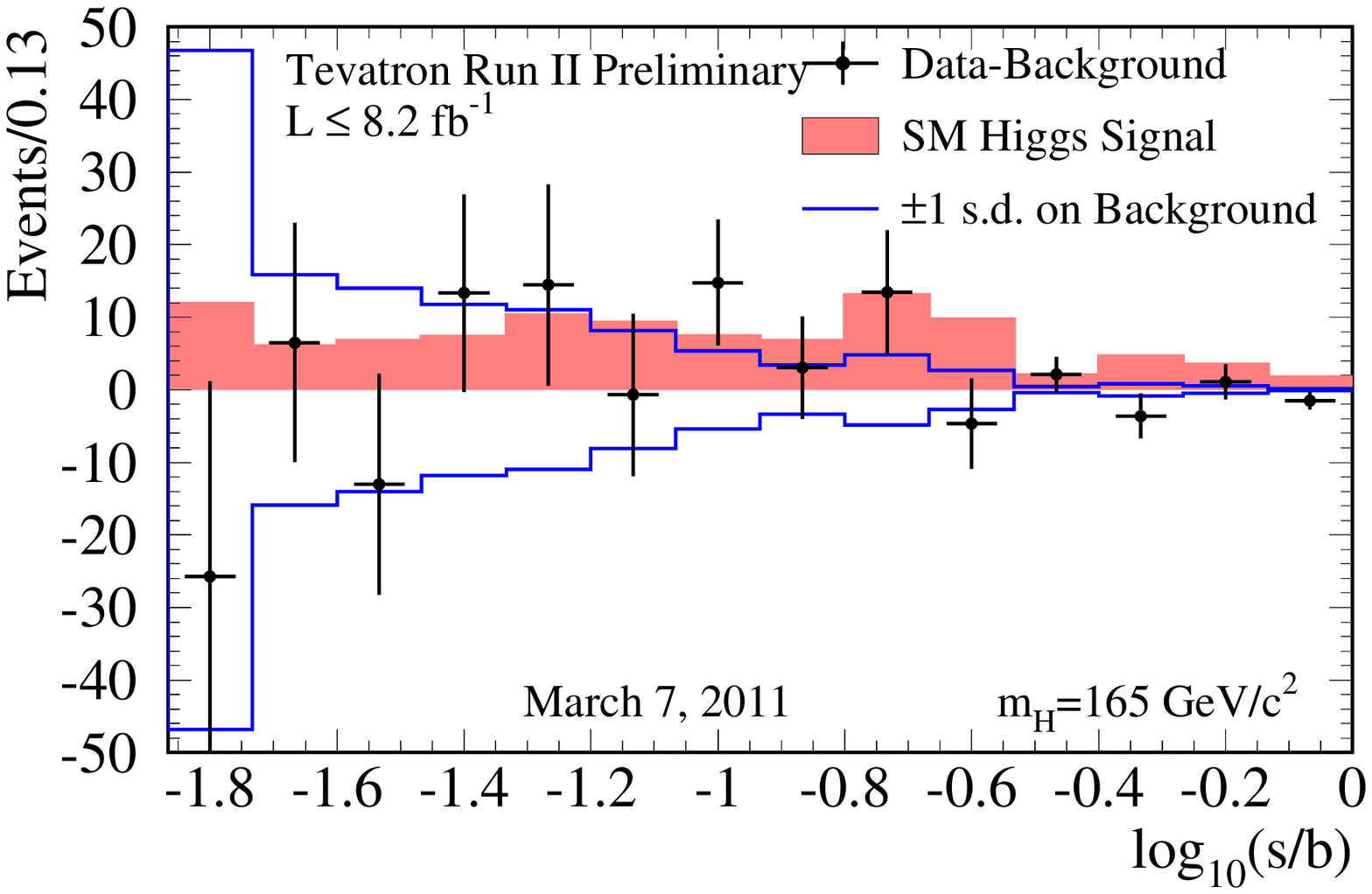}
 \caption{ 
 \label{fig:bgsub} Background-subtracted data distributions for all channels, summed in bins of $s/b$,
for Higgs boson masses of 150, and 165~GeV/$c^2$.  The background has been fit, within its systematic
uncertainties, to the data.  The points with error bars indicate the background-subtracted data; the sizes of the
error bars are the square roots of the predicted background in each bin.  The unshaded (blue-outline) histogram shows the systematic
uncertainty on the best-fit background model, and the shaded histogram shows the expected signal for a Standard
Model Higgs boson.}
 \end{centering}
 \end{figure}

%%%%%%%%%%%%%%%%%%%%%%%%%%%%%%%%%%%%%%%%%%%%%%%%%%%%%%%%%%%%%%%%%%%%%%%
%%%%%%%%%%%%%%%%%%%%%%%%%%%%%%%%%%%%%%%%%%%%%%%%%%%%%%%%%%%%%%%%%%%%%%%
\section{Combining Channels} %%%%%%%%%%%%%%
\label{sec:combination}

To gain confidence that the final result does not depend on the
details of the statistical formulation, we perform two types of
combinations, using Bayesian and Modified Frequentist approaches,
which yield limits on the Higgs boson production rate that agree
within 10\% at each value of $m_H$, and within 1\% on average.  Both
methods rely on distributions in the final discriminants, and not just
on their single integrated values.  Systematic uncertainties enter on
the predicted number of signal and background events as well as on the
distribution of the discriminants in each analysis (``shape
uncertainties'').  Both methods use likelihood calculations based on
Poisson probabilities.

Both methods treat the systematic uncertainties in a Bayesian fashion,
assigning a prior distribution to each source of uncertainty,
parameterized by a nuisance parameter, and propagating the impacts of
varying each nuisance parameter to the signal and background
predictions, with all correlations included.  A single nuisance
parameter may affect the signal and background predictions in many
bins of many different analyses.  Independent nuisance parameters are
allowed to vary separately within their prior distributions.
Both methods use the data to constrain the values of the nuisance
parameters, one by integration, the other by fitting.  These methods
reduce the impact of prior uncertainty in the nuisance parameters thus
improving the sensitivity.  Because of these constraints to the data,
it is important to evaluate the uncertainties and correlations
properly, and to allow independent parameters to vary separately,
otherwise a fit may overconstrain a parameter and extrapolate its use
improperly.  The impacts of correlated uncertainties add together
linearly on a particular prediction, while those of uncorrelated
uncertainties are convoluted together, which is similar to adding in
quadrature.  Adding uncorrelated uncertainties linearly implies a
correlation which is not present and which may result in incorrect
results.

Both methods set limits at the 95\% confidence (or ``credibility'' in
the case of the Bayesian method).  This is a probabalistic statement.
To be consistent and accurate, we give equal treatment to all sources
of uncertainty, both theoretical and experimental.
Ref.~\cite{bagliocrit} proposes testing the minimum possible
production cross section and quoting our exclusion limits based on
only this prediction.  We do not claim that our exclusion limits are
independent of the model choices that are made, and hence we are not
required to only test the case in which all values of all uncertain
parameters fluctuate simultaneously to the values that produce the
weakest results. To do so would provide inconsistent results when
computing limits and when attempting to discover.  A set of values for
the nuisance parameters which weakens the limit may strengthen a
discovery and vice versa, particularly those parameters that affect
the background rate and shape predictions.  By always setting
uncertain parameters to their most extreme values, we could find we
have an excess of data over the background prediction when we set our
limits, and a deficit of data with respect to the background in the
same sample when attempting to make a discovery.  The prescriptions
described below provide a consistent method for both tasks.

\subsection{Bayesian Method}

Because there is no experimental information on the production cross
section for the Higgs boson, in the Bayesian
technique~\cite{prevhiggs}\cite{pdgstats} we assign a flat prior for
the total number of selected Higgs events.  For a given Higgs boson
mass, the combined likelihood is a product of likelihoods for the
individual channels, each of which is a product over histogram bins:

\begin{equation}
{\cal{L}}(R,{\vec{s}},{\vec{b}}|{\vec{n}},{\vec{\theta}})\times\pi({\vec{\theta}})
%{\cal{L}}(R,\vec s, \vec b |\vec n,{\vec{\theta}})
= \prod_{i=1}^{N_C}\prod_{j=1}^{N_b} \mu_{ij}^{n_{ij}} e^{-\mu_{ij}}/n_{ij}!
\times\prod_{k=1}^{n_{np}}e^{-\theta_k^2/2}
\end{equation}

\noindent where the first product is over the number of channels
($N_C$), and the second product is over $N_b$ histogram bins containing
$n_{ij}$ events, binned in  ranges of the final discriminants used for
individual analyses, such as the dijet mass, neural-network outputs,
or matrix-element likelihoods.
 The parameters that contribute to the
expected bin contents are $\mu_{ij} =R \times s_{ij}({\vec{\theta}}) + b_{ij}({\vec{\theta}})$
for the
channel $i$ and the histogram bin $j$, where $s_{ij}$ and $b_{ij}$
represent the expected background and signal in the bin, and $R$ is a scaling factor
applied to the signal to test the sensitivity level of the experiment.
%The likelihood function also contains
%truncated Gaussian constraints on the nuisance parameters $\theta_k$, which define
Truncated Gaussian priors are used for each of the nuisance parameters
$\theta_k$, which define the sensitivity of the predicted signal and
background estimates to systematic uncertainties.  These can take the
form of uncertainties on overall rates, as well as the shapes of the
distributions used for combination.  These systematic uncertainties
can be far larger than the expected SM Higgs boson signal, and are
therefore important in the calculation of limits.  The truncation is
applied so that no prediction of any signal or background in any bin
is negative.  The posterior density function is then integrated over
all parameters (including correlations) except for $R$, and a 95\%
credibility level upper limit on $R$ is estimated by calculating the
value of $R$ that corresponds to 95\% of the area of the resulting
distribution.

\subsection{Modified Frequentist Method}

The Modified Frequentist technique relies on the ${\rm CL}_{\rm s}$
method, using a log-likelihood ratio (LLR) as test
statistic~\cite{DZHiggs}:
\begin{equation}
LLR = -2\ln\frac{p({\mathrm{data}}|H_1)}{p({\mathrm{data}}|H_0)},
\end{equation}
where $H_1$ denotes the test hypothesis, which admits the presence of
SM backgrounds and a Higgs boson signal, while $H_0$ is the null
hypothesis, for only SM backgrounds.  The probabilities $p$ are
computed using the best-fit values of the nuisance parameters for each
pseudo-experiment, separately for each of the two hypotheses, and
include the Poisson probabilities of observing the data multiplied by
Gaussian priors for the values of the nuisance parameters.  This
technique extends the LEP procedure~\cite{pdgstats} which does not
involve a fit, in order to yield better sensitivity when expected
signals are small and systematic uncertainties on backgrounds are
large~\cite{pflh}.

The ${\rm CL}_{\rm s}$ technique involves computing two $p$-values,
${\rm CL}_{\rm s+b}$ and ${\rm CL}_{\rm b}$.  The latter is defined by
\begin{equation}
1-{\rm CL}_{\rm b} = p(LLR\le LLR_{\mathrm{obs}} | H_0),
\end{equation}
where $LLR_{\mathrm{obs}}$ is the value of the test statistic computed
for the data. $1-{\rm CL}_{\rm b}$ is the probability of observing a
signal-plus-background-like outcome without the presence of signal,
i.e. the probability that an upward fluctuation of the background
provides a signal-plus-background-like response as observed in data.
The other $p$-value is defined by
\begin{equation}
{\rm CL}_{\rm s+b} = p(LLR\ge LLR_{\mathrm{obs}} | H_1),
\end{equation}
and this corresponds to the probability of a downward fluctuation of the sum
of signal and background in
the data.  A small value of ${\rm CL}_{\rm s+b}$ reflects inconsistency with  $H_1$.
It is also possible to have a downward fluctuation in data even in the absence of
any signal, and a small value of ${\rm CL}_{\rm s+b}$ is possible even if the expected signal is
so small that it cannot be tested with the experiment.  To minimize the possibility
of  excluding  a signal to which there is insufficient sensitivity
(an outcome  expected 5\% of the time at the 95\% C.L., for full coverage),
we use the quantity ${\rm CL}_{\rm s}={\rm CL}_{\rm s+b}/{\rm CL}_{\rm b}$.  If ${\rm CL}_{\rm s}<0.05$ for a particular choice
of $H_1$, that hypothesis is deemed to be excluded at the 95\% C.L. In an analogous
way, the expected ${\rm CL}_{\rm b}$, ${\rm CL}_{\rm s+b}$ and ${\rm CL}_{\rm s}$ values are computed from the median of the
LLR distribution for the background-only hypothesis.

Dividing CL$_{\rm s+b}$ by CL$_{\rm b}$ incurs a median penalty of a factor of two in the expected $p$-value,
as the distribution of CL$_{\rm b}$ is uniform under the null hypothesis.  One may use CL$_{\rm s+b}$ directly
to set limits, and solve the problem of excluding non-testable models by introducing a
``power-constraint''.  The power-constrained limit (PCL) method~\cite{atlasww,atlaszz,atlasgammagamma}
involves excluding models
for which CL$_{\rm s+b}<0.05$, but if the limit thus obtained makes an excursion below a previously chosen
lower bound, to quote the lower-bound limit instead.  The ATLAS collaboration sets this lower bound at $-1\sigma$
in the distribution of limits expected from background-only outcomes.  The PCL method thus retains
the desired coverage rate and does not exclude untestable models, and it
provides by construction stronger expected and observed limits than CL$_{\rm s}$.  The expected and observed PCL
limits with Tevatron data may be estimated from Figure~\ref{fig:comboCLSB} and Table~\ref{tab:clsbVals} which show and list the observed and expected values
of CL$_{\rm s+b}$ as functions of $m_H$.  Nowhere in the range is there more than a $1\sigma$ downward fluctuation
relative to the background prediction, and so a similar power constraint to the one ATLAS applies would not have
an effect on the observed limits.  With the PCL method, the region of $m_H$ expected to be excluded at the 95\% C.L.
grows by $\sim$40\% compared to that obtained with the CL$_{\rm s}$ method.
The PCL method has a larger false exclusion rate than the CL$_{\rm s}$ and Bayesian
methods that we use to quote our results, 
the distribution of possible limits under the null hypothesis is highly asymmetrical,
and this distribution depends on the arbitrary choice of the location of the power constraint.
We decided to continue quoting CL$_{\rm s}$ and Bayesian limits also because they provide a strong numerical cross-check
of each other, and they are comparable to those of LEP~\cite{lephiggs}, CMS~\cite{cmsww}, and our own
previous results~\cite{prevhiggs}.

Systematic uncertainties are included  by fluctuating the predictions for
signal and background rates in each bin of each histogram in a correlated way when
generating the pseudo-experiments used to compute ${\rm CL}_{\rm s+b}$ and ${\rm CL}_{\rm b}$.

\subsection{Systematic Uncertainties} %%%%%%%%%%%%%%
\label{systematics}

Systematic uncertainties differ
between experiments and analyses, and they affect the rates and shapes of the predicted
signal and background in correlated ways.  The combined results incorporate
the sensitivity of predictions to  values of nuisance parameters,
and include correlations between rates and shapes, between signals and backgrounds,
and between channels within experiments and between experiments.
More discussion on this topic can be found in the
individual analysis notes~\cite{cdfHWW} through~\cite{dzHgg}.  Here we
consider only the largest contributions and correlations between and
within the two experiments.

\subsubsection{Correlated Systematics between CDF and D0}%%%%%%%%%%%%%%

The uncertainties on the measurements of the integrated luminosities are 6\%
(CDF) and 6.1\% (D0).
Of these values, 4\% arises from the uncertainty
on the inelastic \pp~scattering cross section, which is correlated
between CDF and D0.
CDF and D0 also share the assumed values and uncertainties on the production cross sections
for top-quark processes (\ttbar~and single top) and for electroweak processes
($WW$, $WZ$, and $ZZ$).  In order to provide a consistent combination, the values of these
cross sections assumed in each analysis are brought into agreement.  We use
$\sigma_{t\bar{t}}=7.04^{+0.24}_{-0.36}~{\rm (scale)}\pm 0.14{\rm (PDF)}\pm 0.30{\rm (mass)}$,
following the calculation of Moch and Uwer~\cite{mochuwer}, assuming
a top quark mass $m_t=173.0\pm 1.2$~GeV/$c^2$~\cite{tevtop10},
and using the MSTW2008nnlo PDF set~\cite{mstw2008}.  Other
calculations of $\sigma_{t\bar{t}}$ are similar~\cite{otherttbar}.

For single top, we use the NLL $t$-channel calculation of Kidonakis~\cite{kid1},
which has been updated using the MSTW2008nnlo PDF set~\cite{mstw2008}~\cite{kidprivcomm}.
For the $s$-channel process we use~\cite{kid2}, again based on the MSTW2008nnlo PDF set.
Both of the cross section values below are the sum of the single $t$ and single ${\bar{t}}$
cross sections, and both assume $m_t=173\pm 1.2$ GeV.
\begin{equation}
\sigma_{t-{\rm{chan}}} = 2.10\pm 0.027 {\rm{(scale)}} \pm 0.18 {\rm{(PDF)}}  \pm 0.045 {\rm{(mass)}}  {\rm {pb}}.
\end{equation}
\begin{equation}
\sigma_{s-{\rm{chan}}} = 1.046\pm 0.006 {\rm{(scale)}} \pm 0.059~{\rm{(PDF)}}  \pm 0.030~{\rm{(mass)}}~{\rm {pb}}.
\end{equation}
Other calculations of $\sigma_{\rm{SingleTop}}$ are
similar for our purposes~\cite{harris}.

mcfm~\cite{mcfm} has been used to compute the NLO cross sections for $WW$, $WZ$,
and $ZZ$ production~\cite{dibo}.  Using a scale choice $\mu_0=M_V^2+p_T^2(V)$ and
the MSTW2008 PDF set~\cite{mstw2008}, the cross section for inclusive $W^+W^-$
production is
\begin{equation}
\sigma_{W^+W^-} = 11.34^{+0.56}_{-0.49}~{\rm{(scale)}}~^{+0.35}_{-0.28} {\rm(PDF)} {\rm{pb}}
\end{equation}
and the cross section for inclusive $W^\pm Z$ production is
\begin{equation}
\sigma_{W^\pm Z} = 3.22^{+0.20}_{-0.17}~{\rm{(scale)}}~^{+0.11}_{-0.08}~{\rm(PDF)}~{\rm{pb}}
\end{equation}
For the $Z$, leptonic decays are used in the definition of the cross section, and
we assume both $\gamma$ and $Z$ exchange.  The $W^\pm Z$ cross section quoted above involves the requirement
$75\leq m_{\ell^+\ell^-}\leq 105$~GeV for the leptons from the neutral current
exchange.  The same dilepton invariant mass requirement is applied to both
sets of leptons in determining the $ZZ$ cross section which is
\begin{equation}
\sigma_{ZZ} = 1.20^{+0.05}_{-0.04}~{\rm{(scale)}}~^{+0.04}_{-0.03}~{\rm(PDF)}~{\rm{pb}}
\end{equation}
For the diboson cross section calculations, we use $|\eta_{\ell}|<5$
for all calculations.  Loosening this requirement to include all
leptons leads to $\sim$+0.4\% change in the predictions.  Lowering the
factorization and renormalization scales by a factor of two increases
the cross section, and raising the scales by a factor of two decreases
the cross section.  The PDF uncertainty has the same fractional impact
on the predicted cross section independent of the scale choice.  All
PDF uncertainties are computed as the quadrature sum of the twenty
68\% C.L. eigenvectors provided with MSTW2008 (MSTW2008nlo68cl).  We
furthermore constrain $\sigma_{W^+W^-}$ in the signal regions of the
channels in the process of setting our limits, either by integration
over the uncertain parameters, or by a direct fit, depending on the
method.  Our posterior constraint on $\sigma_{W^+W^-}$ has a
fractional uncertainty of $\pm 4$\%, which includes the prior
theoretical constraint of $\pm 6$\%, indicating that the data and the
theoretical prediction are approximately equally constraining on
$\sigma_{W^+W^-}$.

In many analyses, the dominant background yields are calibrated with data control
samples.  Since the methods of measuring the multijet (``QCD'') backgrounds differ
between CDF and D0, and even between analyses within the collaborations, there is
no correlation assumed between these rates.  Similarly, the large uncertainties on
the background rates for $W$+heavy flavor (HF) and $Z$+heavy flavor are considered
at this time to be uncorrelated, as both CDF and D0 estimate these rates using data
control samples, but employ different techniques.  The calibrations of fake leptons,
unvetoed $\gamma\rightarrow e^+e^-$ conversions, $b$-tag efficiencies and mistag
rates are performed by each collaboration using independent data samples and
methods, and are therefore also treated as uncorrelated.

\subsubsection{Correlated Systematic Uncertainties for CDF}%%%%%%%%%%%%%%
The dominant systematic uncertainties for the CDF analyses are shown
in the Appendix
in Tables~\ref{tab:cdfsystww0}, \ref{tab:cdfsystww4}, and~\ref{tab:cdfsystww5}
for the $H \rightarrow W^+W^-\rightarrow \ell^{\prime \pm}\nu \ell^{\prime\mp}\nu$ channels,
in Table~\ref{tab:cdfsystwww} for the $WH \rightarrow WWW \rightarrow\ell^{\prime \pm}\ell^{\prime \pm}$ and
$WH\rightarrow WWW \rightarrow \ell^{\pm}\ell^{\prime \pm} \ell^{\prime \prime \mp}$ channels, and
in Table~\ref{tab:cdfsystzww} for the $ZH \rightarrow ZWW \rightarrow \ell^{\pm}\ell^{\mp} \ell^{\prime \pm}$ channels.
Each source induces a
correlated uncertainty across all CDF channels' signal and background
contributions which are sensitive to that source.
Shape dependencies of templates on jet energy scale
and ``FSR'') are taken into account in the analyses (see tables).
For \hww, the largest uncertainties on signal acceptance originate from
Monte Carlo modeling.  Uncertainties on background event rates vary
significantly for the different processes.  The backgrounds with the
largest systematic uncertainties are in general quite small. Such
uncertainties are constrained by fits to the nuisance parameters, and
they do not affect the result significantly.  Because the largest
background contributions are measured using data, these uncertainties
are treated as uncorrelated for the \hbb~channels.  The differences in
the resulting limits when treating the remaining uncertainties as either
correlated or uncorrelated, is less than $5\%$.

\subsubsection{Correlated Systematic Uncertainties for D0 }%%%%%%%%%%%%%%
The dominant systematic uncertainties for the D0 analyses are shown in the Appendix, in
Tables~\ref{tab:d0systww}, \ref{tab:d0systwwtau}, \ref{tab:d0systwww}, \ref{tab:d0lvjj},
\ref{tab:d0sysVHtau}, and \ref{tab:d0systgg}.  Each source induces a correlated uncertainty
across all D0 channels sensitive to that source. Wherever appropriate the impact of systematic
effects on both the rate and shape of the predicted signal and background is included.  
For the \hww and \vww analyses, a significant source
of uncertainty is the measured efficiencies for selecting leptons.  Significant sources for
all analyses are the uncertainties on the luminosity and the cross sections for the simulated
backgrounds. 
For analyses involving jets the determination of the jet energy scale, jet
resolution and
the multijet background contribution are significant sources of uncertainty.
All systematic uncertainties arising from the same source are taken to be
correlated among the different backgrounds and between signal and background.

 \begin{figure}[t]
 \begin{centering}
 \includegraphics[width=14.0cm]{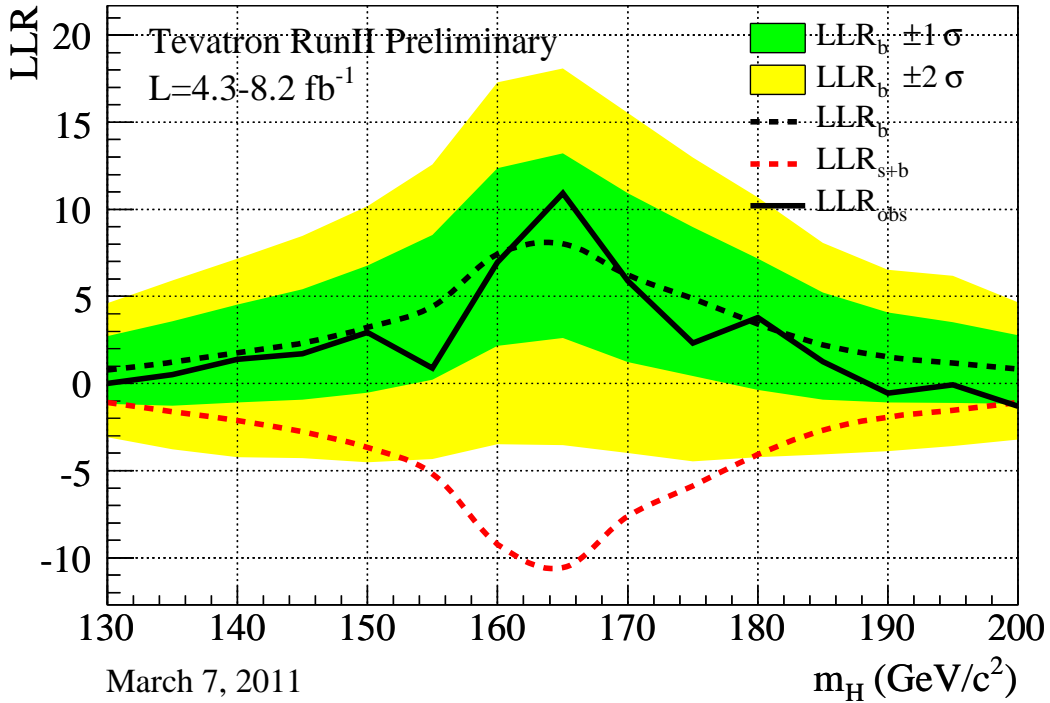}
 \caption{
 \label{fig:comboLLR} {
Distributions of the log-likelihood ratio (LLR) as a function of Higgs mass obtained with the ${\rm CL}_{\rm s}$ method for the combination of all CDF and D0 analyses.
}}
 \end{centering}
 \end{figure}

\vspace*{1cm}
%%%%%%%%%%%%%%%%%%%%%%%%%%%%%%%%%%%%%%%%%%%%%%%%%%%%%%%%%%%%%%%%%%%%%%%
%%%%%%%%%%%%%%%%%%%%%%%%%%%%%%%%%%%%%%%%%%%%%%%%%%%%%%%%%%%%%%%%%%%%%%%
\section{Combined Results} %%%%%%%%%%%%%%

Before extracting the combined limits we study the distributions of the
log-likelihood ratio (LLR) for different hypotheses, to quantify the expected
sensitivity across the mass range tested.
Figure~\ref{fig:comboLLR} displays the LLR distributions for the combined
analyses as functions of $m_{H}$. Included are the median of the LLR distributions for the
background-only hypothesis (LLR$_{b}$), the signal-plus-background
hypothesis (LLR$_{s+b}$), and the observed value for the data (LLR$_{\rm{obs}}$).  The
shaded bands represent the one and two standard deviation ($\sigma$)
departures for LLR$_{b}$ centered on the median. Table~\ref{tab:llrVals} lists the observed
and expected LLR values shown in Figure~\ref{fig:comboLLR}.

%%%%%%%%%%%%%%%%%%%%%%%%%%%%%%%%%%%%%%%%%%%%%%%%%%%%%%%%%%%%%%%%%%%%%
\begin{table}[htpb]
\caption{\label{tab:llrVals}
 Log-likelihood ratio (LLR) values for the combined CDF + \Dzero Higgs boson search obtained using the {\rm CL}$_{S}$ method.}
\begin{ruledtabular}
\begin{tabular}{lccccccc}
$m_{H}$ (GeV/$c^2$ &  LLR$_{\rm{obs}}$ & LLR$_{S+B}^{\rm{med}}$ &
LLR$_{B}^{-2\sigma}$ & LLR$_{B}^{-1\sigma}$ & LLR$_{B}^{\rm{med}}$ &  LLR$_{B}^{+1\sigma}$ & LLR$_{B}^{+2\sigma}$ \\
\hline
130 & 0.01 & -1.07 & 4.62 & 2.73 & 0.78 & -1.18 & -3.08 \\
135 & 0.52 & -1.62 & 5.92 & 3.58 & 1.23 & -1.27 & -3.77 \\
140 & 1.39 & -2.12 & 7.17 & 4.53 & 1.77 & -1.07 & -4.22 \\
145 & 1.71 & -2.77 & 8.47 & 5.42 & 2.33 & -0.93 & -4.28 \\
150 & 2.95 & -3.67 & 10.18 & 6.78 & 3.23 & -0.53 & -4.53 \\
155 & 0.89 & -5.17 & 12.57 & 8.53 & 4.42 & 0.23 & -4.33 \\
160 & 6.96 & -9.22 & 17.27 & 12.38 & 7.42 & 2.17 & -3.48 \\
165 & 10.93 & -10.57 & 18.07 & 13.22 & 8.03 & 2.62 & -3.52 \\
170 & 5.90 & -7.62 & 15.53 & 10.93 & 6.22 & 1.23 & -3.98 \\
175 & 2.33 & -5.88 & 12.97 & 8.97 & 4.88 & 0.42 & -4.47 \\
180 & 3.78 & -4.08 & 10.68 & 7.17 & 3.42 & -0.38 & -4.22 \\
185 & 1.26 & -2.67 & 8.07 & 5.22 & 2.23 & -0.93 & -4.08 \\
190 & -0.56 & -1.93 & 6.53 & 4.08 & 1.52 & -1.07 & -3.88 \\
195 & -0.08 & -1.52 & 6.17 & 3.52 & 1.18 & -1.12 & -3.58 \\
200 & -1.30 & -1.07 & 4.67 & 2.77 & 0.82 & -1.18 & -3.23 \\ 
\end{tabular}
\end{ruledtabular}
\end{table}

These distributions can be interpreted as follows: The separation
between the medians of the LLR$_{b}$ and LLR$_{s+b}$ distributions
provides a measure of the discriminating power of the search.  The
sizes of the one- and two-$\sigma$ LLR$_{b}$ bands indicate the width
of the LLR$_{b}$ distribution, assuming no signal is truly present and
only statistical fluctuations and systematic effects are present.  The
value of LLR$_{\rm{obs}}$ relative to LLR$_{s+b}$ and LLR$_{b}$
indicates whether the data distribution appears to resemble what we
expect if a signal is present (i.e. closer to the LLR$_{s+b}$
distribution, which is negative by construction) or whether it
resembles the background expectation more closely; the significance of
any departures of LLR$_{\rm{obs}}$ from LLR$_{b}$ can be evaluated by
the width of the LLR$_{b}$ bands.

Using the combination procedures outlined in Section III, we extract
limits on SM Higgs boson production $\sigma \times B(H\rightarrow X)$
in \pp~collisions at $\sqrt{s}=1.96$~TeV for $130\leq m_H \leq 200$
GeV/$c^2$.  To facilitate comparisons with the SM and to
accommodate analyses with different degrees of sensitivity, we present
our results in terms of the ratio of obtained limits to the SM Higgs
boson production cross section, as a function of Higgs boson mass, for
test masses for which both experiments have performed dedicated
searches in different channels.  A value of the combined limit ratio
which is equal to or less than one indicates that that particular
Higgs boson mass is excluded at the 95\% C.L.  A value less than one indicates
that a Higgs boson of that mass is excluded with a smaller cross section than
the SM prediction, and that the SM prediction is excluded with more
certainty than 95\% C.L.

The combinations of results~\cite{cdfHWW,DZHiggs} of each single
experiment, as used in this Tevatron combination, yield the following
ratios of 95\% C.L.  observed (expected) limits to the SM cross
section: 0.92~(0.93) for CDF and 0.75~(0.92) for D0 at
$m_{H}=165$~GeV/$c^2$.  Both collaborations independently exclude
$m_{H}=165$~GeV/$c^2$ at the 95\% C.L.

The ratios of the 95\% C.L. expected and observed limit to the SM
cross section are shown in Figure~\ref{fig:comboRatio} for the
combined CDF and D0 analyses.  The observed and median expected ratios
are listed for the tested Higgs boson masses in Table~\ref{tab:ratios}
for $m_{H}
\leq 150$~GeV/$c^2$, and in Table~\ref{tab:ratios-3} for $m_{H} \geq
155$~GeV/$c^2$, as obtained by the Bayesian and the ${\rm CL}_{\rm s}$
methods.  In the following summary we quote only the limits obtained
with the Bayesian method, which was chosen {\it a priori}.  It
turns out that the Bayesian limits are slightly less stringent.  The
corresponding limits and expected limits obtained using the ${\rm
CL}_{\rm s}$ method are shown alongside the Bayesian limits in the
tables.  We obtain the observed (expected) values of 1.31~(0.92) at
$m_{H}=155$~GeV/$c^2$, 0.54~(0.65) at $m_{H}=165$~GeV/$c^2$,
1.13~(0.85) at $m_{H}=175$~GeV/$c^2$ and 1.49~(1.30) at
$m_{H}=185$~GeV/$c^2$.

\begin{figure}[hb]
\begin{centering}
\includegraphics[width=16.5cm]{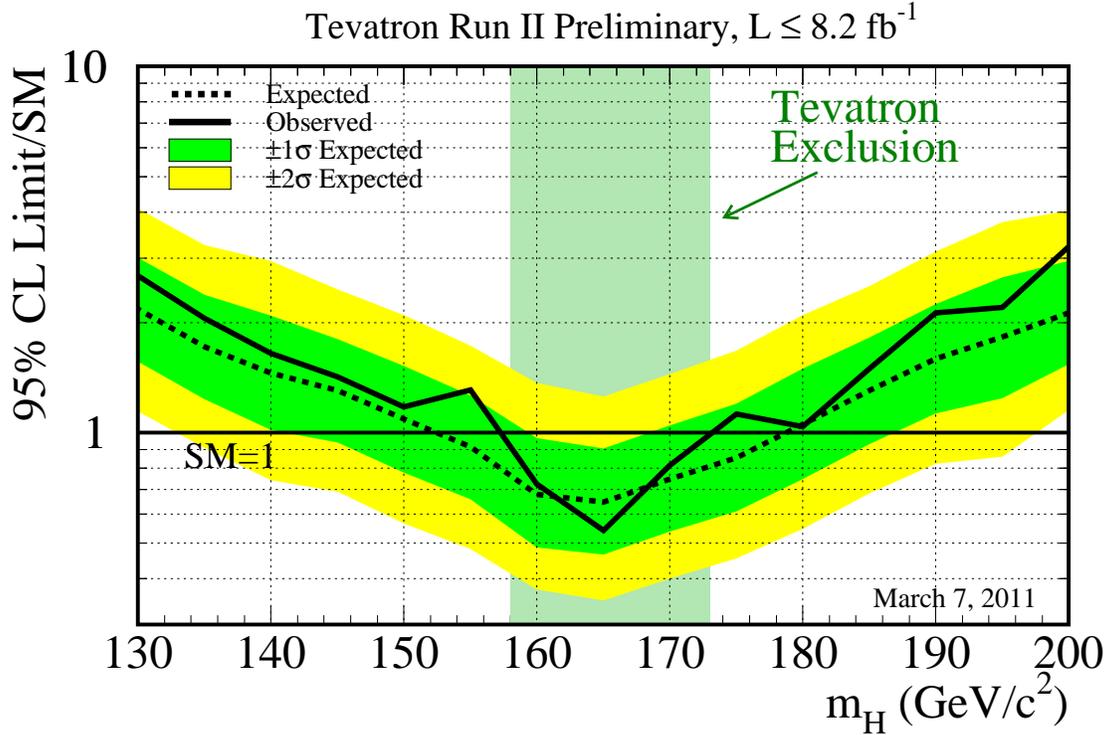}
\caption{
\label{fig:comboRatio} 
Observed and expected (median, for the background-only hypothesis)
95\% C.L. upper limits on the ratios to the SM cross section, as
functions of the Higgs boson mass for the combined CDF and D0
analyses.  The limits are expressed as a multiple of the SM prediction
for test masses (every 5 GeV/$c^2$) for which both experiments have
performed dedicated searches in different channels.  The points are
joined by straight lines for better readability.  The bands indicate
the 68\% and 95\% probability regions where the limits can fluctuate,
in the absence of signal.  The limits displayed in this figure are
obtained with the Bayesian calculation.  }
\end{centering}
\end{figure}

\begin{table}[ht]
\caption{\label{tab:ratios}  Ratios of median expected and observed 95\% C.L.
limit to the SM cross section for the combined CDF and D0 analyses as a function
of the Higgs boson mass in GeV/$c^2$, obtained with the Bayesian and with the ${\rm CL}_{\rm s}$ method.}
\begin{ruledtabular}
\begin{tabular}{lccccc}\\
Bayesian       &  130 &  135 &  140 &  145 &  150 \\ \hline
Expected       & 2.18 & 1.72 & 1.46 & 1.31 & 1.09 \\
Observed       & 2.69 & 2.05 & 1.65 & 1.42 & 1.18 \\

\hline
\hline\\
${\rm CL}_{\rm s}$         &  130 &  135 &  140 &  145 &  150 \\ \hline
Expected:                  &2.14  &1.72  &1.49  &1.29  &1.14 \\
Observed:                  &2.57  &1.98  &1.60  &1.42  &1.16 \\
\end{tabular}
\end{ruledtabular}
\end{table}

\begin{table}[ht]
\caption{\label{tab:ratios-3}
Ratios of median expected and observed 95\% C.L.
limit to the SM cross section for the combined CDF and D0 analyses as a function
of the Higgs boson mass in GeV/$c^2$, obtained with the Bayesian and with the ${\rm CL}_{\rm s}$ method.}
\begin{ruledtabular}
\begin{tabular}{lccccccccccc}
Bayesian             &  155 &  160 &  165 &  170 &  175 &  180 &  185 &  190 &  195 &  200 \\ \hline
Expected             & 0.92 & 0.68 & 0.65 & 0.75 & 0.85 & 1.06 & 1.30 & 1.59 & 1.83 & 2.13 \\
Observed             & 1.31 & 0.72 & 0.54 & 0.82 & 1.13 & 1.04 & 1.49 & 2.13 & 2.20 & 3.22 \\
\hline
\hline\\
${\rm CL}_{\rm s}$   &  155 &  160 &  165 &  170 &  175 &  180 &  185 &  190 &  195 &  200 \\ \hline
Expected             & 0.92 &0.68 &0.64 &0.77 &0.87 &1.05 &1.32 &1.60 &1.82 &2.09 \\
Observed             & 1.28 &0.70 &0.52 &0.80 &1.09 &1.03 &1.49 &2.13 &2.22 &3.13 \\
\end{tabular}
\end{ruledtabular}
\end{table}

We show in Figure~\ref{fig:comboCLS} and list in
Table~\ref{tab:clsVals} the observed 1-${\rm CL}_{\rm s}$ and its
expected distribution for the background-only hypothesis as a function
of the Higgs boson mass. This is directly interpreted as the level of
exclusion of our search using the ${\rm CL}_{\rm s}$ method.  The
region excluded at the 95\% C.L. agrees very well with that obtained
via the Bayesian calculation.

In addition, we provide in Figure~\ref{fig:comboCLSB} (and listed in
Table~\ref{tab:clsbVals}) the values for the observed 1-${\rm CL}_{\rm s+b}$ 
and its expected distribution as a function of $m_H$.  The value
${\rm CL}_{\rm s+b}$ is the $p$-value for the signal-plus-background
hypothesis.  These values can be used to obtain alternative
upper limits that are more constraining compared to those obtained via the ${\rm CL}_{\rm s}$ 
method.  In such a formulation, the power of the search is limited
at a level chosen {\it a priori} to avoid setting limits when the
background model grossly overpredicts the data or the data exhibit a
large background-like fluctuation ({\it e.g.}, limit at the -1$\sigma$
background fluctuation level.).  Within Figure~\ref{fig:comboCLSB}, 95\%
C.L.  power-constrained limits can be found at the points for which
1-${\rm CL}_{\rm s+b}$ exceeds 95\%.  The expected range of exclusion is
$\sim$40\% larger using PCL than the Bayesian and CL$_{\rm s}$ limits quoted
here.  We continue our convention of quoting Bayesian and CL$_{\rm s}$ limits
however.

In summary, we combine CDF and D0 results on SM Higgs boson searches,
based on luminosities up to 8.2 fb$^{-1}$.
Compared to our previous combination, more data have been added to the existing
channels, additional channels have been included, and analyses have been further
optimized to gain sensitivity. We use the recommendation of the PDF4LHC
working group for the central value of the parton distribution functions and uncertainties~\cite{pdf4lhc}.
We use the highest-order calculations of $gg \rightarrow H$, $WH$, $ZH$, and VBF theoretical cross 
sections when comparing our limits to the SM predictions at high mass.  We include consensus estimates
of the theoretical uncertainties on these production cross sections and the decay
branching fractions in the computations of our limits.

The 95\% C.L. upper limit on Higgs boson production is a factor of 0.54
times the SM cross section for a Higgs boson mass of $m_{H}=$165~GeV/$c^2$.
Based on simulation, the corresponding median expected upper limit is 0.65 times the SM cross section.
Standard Model branching ratios, calculated as functions of the Higgs boson mass, are assumed.

We choose to use the intersections of piecewise linear interpolations of our observed and expected
rate limits in order to quote ranges of Higgs boson masses that are excluded and that are expected
to be excluded.  The sensitivities of our searches to Higgs bosons are smooth functions of the Higgs
boson mass and depend most strongly on the predicted cross sections and the decay branching ratios
(the decay $H\rightarrow W^+W^-$ is the dominant decay for the region of highest sensitivity). The
mass resolution of the channels is poor due to the presence of two highly energetic neutrinos in
signal events.  We therefore use the linear interpolations to extend the results from the 5~GeV/$c^2$
mass grid investigated to points in between.  This procedure yields higher expected and observed
interpolated rate limits than if the full dependence of the cross section and branching ratio were
included as well, since the latter produces limit curves that are concave upwards.
We exclude in this way the region $158<m_{H}<173$~GeV/$c^{2}$ at the the 95\% C.L.
The expected exclusion region, given the current sensitivity, is
$153<m_{H}<179$~GeV/$c^{2}$.  The excluded region obtained by finding the intersections of the
linear interpolations of the observed $1-{\rm CL}_{\rm s}$ curve shown in Figure~\ref{fig:comboCLS} is
slightly larger than that obtained with the Bayesian calculation.  As previously stated, and following
the procedure used in previous combinations~\cite{prevhiggs}, we make
the {\it a priori} choice to quote the exclusion region using the Bayesian calculation.

The results presented in this paper significantly extend the individual limits of each
collaboration and those obtained in our previous combination.  The sensitivity of our combined search is
sufficient to exclude a Higgs boson at high mass and is expected to grow substantially
in the future as more data are added and further improvements are made to the analysis
techniques.

 \begin{figure}[t]
 \begin{centering}
 \includegraphics[width=14.0cm]{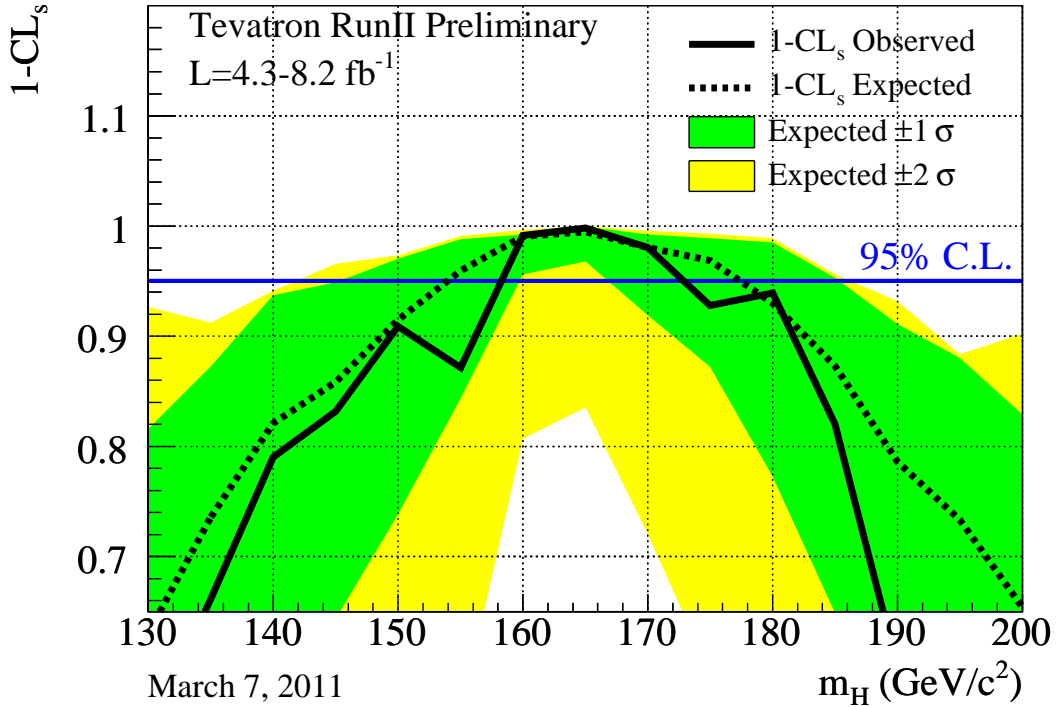}
 \caption{
 \label{fig:comboCLS}
 The exclusion strength 1-${\rm CL}_{\rm s}$ as a function of the Higgs boson mass
(in steps of 5 GeV/$c^2$), as obtained with ${\rm CL}_{\rm s}$ method
 %obtained with the ${\rm CL}_{\rm s}$ method
 for the combination of the
 CDF and D0 analyses. }
 \end{centering}
 \end{figure}

%%%%%%%%%%%%%%%%%%%%%%%%%%%%%%%%%%%%%%%%%%%%%%%%%%%%%%%%%%%%%%%%%%%%%%%

\begin{table}[htpb]
\caption{\label{tab:clsVals}
The observed and expected 1-{\rm CL}$_{\rm s}$ values as functions of $m_H$, for the combined
CDF and \Dzero Higgs boson searches.}
\begin{ruledtabular}
\begin{tabular}{lcccccc}
$m_H$ (GeV/$c^2$) & 1-{\rm CL}$_{\rm s}^{\rm{obs}}$ &
1-{\rm CL}$_{\rm s}^{-2\sigma}$ &
1-{\rm CL}$_{\rm s}^{-1\sigma}$ &
1-{\rm CL}$_{\rm s}^{\rm{median}}$ &
1-{\rm CL}$_{\rm s}^{+1\sigma}$ &
1-{\rm CL}$_{\rm s}^{+2\sigma}$ \\ \hline
130 & 0.530 & 0.929 & 0.816 & 0.638 & 0.366 & 0.130\\ 
135 & 0.654 & 0.925 & 0.888 & 0.729 & 0.466 & 0.167\\ 
140 & 0.790 & 0.923 & 0.919 & 0.811 & 0.549 & 0.233\\ 
145 & 0.840 & 0.956 & 0.950 & 0.865 & 0.636 & 0.283\\ 
150 & 0.909 & 0.985 & 0.980 & 0.916 & 0.737 & 0.398\\ 
155 & 0.864 & 0.987 & 0.984 & 0.951 & 0.834 & 0.545\\ 
160 & 0.991 & 0.995 & 0.992 & 0.990 & 0.952 & 0.795\\ 
165 & 0.998 & 1.000 & 0.999 & 0.995 & 0.967 & 0.839\\ 
170 & 0.982 & 0.997 & 0.994 & 0.987 & 0.923 & 0.727\\ 
175 & 0.933 & 0.999 & 0.995 & 0.970 & 0.876 & 0.614\\ 
180 & 0.941 & 0.989 & 0.980 & 0.929 & 0.785 & 0.476\\
185 & 0.820 & 0.947 & 0.943 & 0.866 & 0.650 & 0.311\\
190 & 0.598 & 0.918 & 0.914 & 0.775 & 0.519 & 0.227\\ 
195 & 0.602 & 0.931 & 0.899 & 0.751 & 0.469 & 0.203\\ 
200 & 0.370 & 0.925 & 0.823 & 0.647 & 0.387 & 0.121\\ 
\end{tabular}
\end{ruledtabular}
\end{table}

 \begin{figure}[t]
 \begin{centering}
 \includegraphics[width=14.0cm]{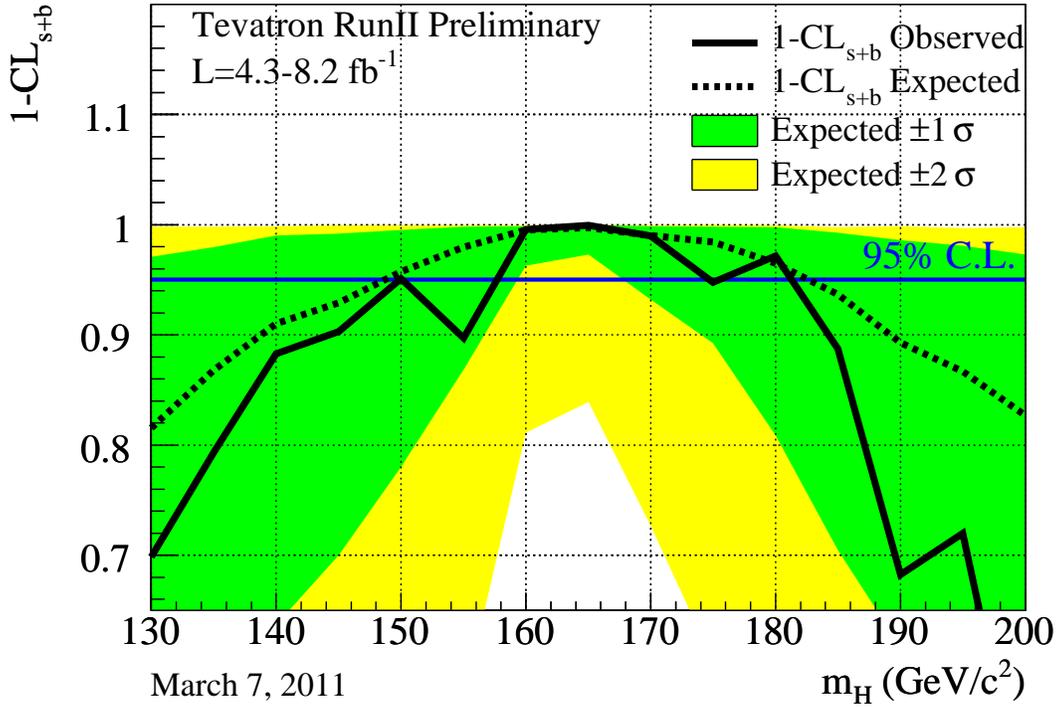}
 \caption{
 \label{fig:comboCLSB}
 The signal exclusion $p$-value 1-${\rm CL}_{\rm s+b}$ as a function of the Higgs boson mass
(in steps of 5 GeV/$c^2$) for the combination of the
 CDF and D0 analyses. }
 \end{centering}
 \end{figure}

\begin{table}[htpb]
\caption{\label{tab:clsbVals}
The observed and expected 1-{\rm CL}$_{\rm s+b}$ values as functions of $m_H$, for the combined CDF and \Dzero Higgs boson searches.}
\begin{ruledtabular}
\begin{tabular}{lcccccc}
$m_H$ (GeV/$c^2$) & 1-{\rm CL}$_{\rm s+b}^{\rm{obs}}$ &
1-{\rm CL}$_{\rm s+b}^{-2\sigma}$ &
1-{\rm CL}$_{\rm s+b}^{-1\sigma}$ &
1-{\rm CL}$_{\rm s+b}^{\rm{median}}$ &
1-{\rm CL}$_{\rm s+b}^{+1\sigma}$ &
1-{\rm CL}$_{\rm s+b}^{+2\sigma}$ \\ \hline
130 & 0.69856 & 0.99833 & 0.97078 & 0.81544 & 0.47078 & 0.16344\\
135 & 0.79378 & 0.99800 & 0.97978 & 0.86800 & 0.55056 & 0.20189\\
140 & 0.88293 & 0.99840 & 0.99000 & 0.91040 & 0.63680 & 0.23773\\
145 & 0.90322 & 0.99922 & 0.99189 & 0.92922 & 0.70000 & 0.32522\\
150 & 0.95089 & 0.99922 & 0.99533 & 0.95744 & 0.78011 & 0.40944\\
155 & 0.89711 & 0.99956 & 0.99811 & 0.97967 & 0.86833 & 0.56667\\
160 & 0.99551 & 0.99956 & 0.99878 & 0.99533 & 0.96289 & 0.81144\\
165 & 0.99951 & 0.99993 & 0.99993 & 0.99746 & 0.97289 & 0.83888\\
170 & 0.98971 & 0.99956 & 0.99878 & 0.99022 & 0.93189 & 0.72667\\
175 & 0.94778 & 0.99967 & 0.99822 & 0.98444 & 0.89267 & 0.60356\\
180 & 0.97163 & 0.99928 & 0.99763 & 0.96493 & 0.80864 & 0.48246\\
185 & 0.88767 & 0.99767 & 0.99256 & 0.93656 & 0.70467 & 0.34278\\
190 & 0.68256 & 0.99844 & 0.98589 & 0.89300 & 0.61711 & 0.24367\\
195 & 0.71978 & 0.99700 & 0.98100 & 0.86678 & 0.55878 & 0.20411\\
200 & 0.45800 & 0.99778 & 0.97289 & 0.82689 & 0.47867 & 0.15800\\
\end{tabular}
\end{ruledtabular}
\end{table}

\clearpage
\newpage

%%%%%%% Test Appendices %%%%%%%%

\appendix
\appendixpage
\addappheadtotoc
\section{Systematic Uncertainties}

%%%% CDF HWW

\begin{table}
\begin{center}
\caption{\label{tab:cdfsystww0} Systematic uncertainties on the signal and background contributions for CDF's
$H\rightarrow W^+W^-\rightarrow\ell^{\pm}\ell^{\prime \mp}$ channels with zero, one, and two or more associated
jets.  These channels are sensitive to gluon fusion production (all channels) and $WH, ZH$ and VBF production.
Systematic uncertainties are listed by name (see the original references for a detailed explanation of their
meaning and on how they are derived).  Systematic uncertainties for $H$ shown in this table are obtained for
$m_H=160$ GeV/$c^2$.  Uncertainties are relative, in percent, and are symmetric unless otherwise indicated.
The uncertainties associated with the different background and signal processed are correlated within individual
jet categories unless otherwise noted.  Boldface and italics indicate groups of uncertainties which are correlated
with each other but not the others on the line.}

\vskip 0.1cm
{\centerline{CDF: $H\rightarrow W^+W^-\rightarrow\ell^{\pm}\ell^{\prime \mp}$ with no associated jet channel relative uncertainties (\%)}}
\vskip 0.099cm
\begin{ruledtabular}
\begin{tabular}{lccccccccccc} \\
Contribution               &   $WW$     &  $WZ$         &  $ZZ$  &  $t\bar{t}$   &  DY    &  $W\gamma$   & $W$+jet &$gg\to H$&  $WH$ &  $ZH$  &  VBF  \\ \hline
{\bf Cross Section :}      &        	&        	&        	&        &        	&        &         &        &        &        & 	\\
Scale                      &        	&        	&        	&        &        	&        &         &  7.0  &        &        &       	\\
PDF Model                  &        	&        	&        	&        &        	&        &         &  7.6  &        &        &       	\\
Total                      & {\it 6.0 }& {\it 6.0 }   & {\it 6.0 }   & 7.0  &  		& 	 & 	   & 	    &  {\bf 5.0 } &  {\bf 5.0 } &       10.0           \\
{\bf Acceptance :}         &        	&        	&        	&        &        	&        &         &        &        &        & 	\\
Scale (jets)               & {\it 0.3 }&        	&        	&        &        	&        &         &       &        &        &      	\\
PDF Model (leptons)        &            &       	&        	&        &              &  	 &         &  2.7  &        &        &         \\
PDF Model (jets)           & {\it 1.1 } &        	&        	&        &        	&        &         &  5.5  &        &        &      	\\
Higher-order Diagrams      &            & {\it 10.0 }  & {\it 10.0 } 	& 10.0  &              & 10.0  &   	   &        & {\bf 10.0 } & {\bf 10.0 } & {\bf 10.0 }          \\
$\MET$ Modeling            &  	 	&      	        &     	        &        & 19.5  	&	 &         &        &        &        &       	\\
Conversion Modeling       &             &               &               &        &              & 10.0  &         &        &        &        &       	\\
Jet Fake Rates             &        	&        	&        	&        &        	&        &         &        &        &        &       	\\
(Low S/B)                  &        	&        	&        	&        &        	&        & 22.0   &        &        &        &       	\\
(High S/B)                 &        	&        	&        	&        &        	&        & 26.0   &        &        &        &       	\\
Jet Energy Scale          &  {\it 2.6 }  & {\it 6.1 }   &   {\it 3.4 }   &   {\it 26.0 }  &  {\it 17.5 } &   {\it 3.1 } &   & {\it 5.0 } & {\it 10.5 } & {\it 5.0 } & {\it 11.5 } \\
Lepton ID Efficiencies     & {\it 3.8 } & {\it 3.8 } 	& {\it 3.8 } 	&{\it 3.8 }&{\it 3.8 }&  	 &         & {\it 3.8 } &  {\it 3.8 } &  {\it 3.8 } &  {\it 3.8 } \\
Trigger Efficiencies       & {\it 2.0 } & {\it 2.0 } 	& {\it 2.0 } 	&{\it 2.0 }&{\it 2.0 }&  	 &         & {\it 2.0 } &  {\it 2.0 } &  {\it 2.0 } &  {\it 2.0 } \\
{\bf Luminosity}           & {\it 3.8 } & {\it 3.8 } 	& {\it 3.8 } 	&{\it 3.8 }&{\it 3.8 }&  	 &         & {\it 3.8 } &  {\it 3.8 } &  {\it 3.8 } &  {\it 3.8 } \\
{\bf Luminosity Monitor}   & {\it 4.4 } & {\it 4.4 } 	& {\it 4.4 } 	&{\it 4.4 }&{\it 4.4 }&  	 &         & {\it 4.4 } &  {\it 4.4 } &  {\it 4.4 } &  {\it 4.4 } \\
\end{tabular}
\end{ruledtabular}

\vskip 0.3cm
{\centerline{CDF: $H\rightarrow W^+W^-\rightarrow\ell^{\pm}\ell^{\prime \mp}$ with one associated jet channel relative uncertainties (\%)}}
\vskip 0.099cm
\begin{ruledtabular}
\begin{tabular}{lccccccccccc} \\
Contribution               &   $WW$     &  $WZ$         &  $ZZ$  &  $t\bar{t}$   &  DY     & $W\gamma$   & $W$+jet &$gg \to H$&  $WH$  &  $ZH$  &  VBF \\ \hline
{\bf Cross Section :}      &        	&        	&        	&        &        	&        &         &        &        &        &        \\
Scale                      &        	&        	&        	&        &        	&        &         & 23.5  &        &        &        \\
PDF Model                  &        	&        	&        	&        &        	&        &         & 17.3  &        &        &        \\
Total                      & {\it 6.0 }& {\it 6.0 }   & {\it 6.0 }   & 7.0   &  	 	& 	 & 	   & 	    &{\bf 5.0 }&{\bf 5.0 }& 10.0  \\
{\bf Acceptance :}         &        	&        	&        	&        &        	&        &         &        &        &        &        \\
Scale (jets)               &{\it -4.0 } &        	&        	&        &        	&        &         &        &        &        &        \\
PDF Model (leptons)        &            &               &               &        &              &  	 &	   &{\it 3.6 }&     &        &        \\
PDF Model (jets)           &{\it  4.7 }&        	&        	&        &        	&        &         & -6.3  &        &        &        \\
Higher-order Diagrams      &            & {\it 10.0 }  & {\it 10.0 }  & 10.0  &  	        & 10.0  &   	   &        &{\bf 10.0 }&{\bf 10.0 }&{\bf 10.0 }\\
$\MET$ Modeling            & 	 	&   	        &       	&        & 20.0  	& 	 &         &        &        &        &        \\
Conversion Modeling       &            &               &               &        &              & 10.0  &         &        &        &        &        \\
Jet Fake Rates             &        	&        	&        	&        &        	&        &         &        &        &        &        \\
(Low S/B)                  &        	&        	&        	&        &        	&        & 23.0   &        &        &        &        \\
(High S/B)                 &        	&        	&        	&        &        	&        & 29.0   &        &        &        &        \\
Jet Energy Scale &  {\it -5.5 }  & {\it -1.0 } &   {\it -4.3 }  &   {\it -13.0 } & {\it -6.5 } &   {\it -9.5 } &   & {\it -4.0 } & {\it -8.5 } & {\it -7.0 } & {\it -6.5 } \\
Lepton ID Efficiencies     & {\it 3.8 } & {\it 3.8 } 	& {\it 3.8 } 	&{\it 3.8 }&{\it 3.8 }& 	 &      &{\it 3.8 }&{\it 3.8 }&{\it 3.8 }&{\it 3.8 }\\
Trigger Efficiencies       & {\it 2.0 } & {\it 2.0 } 	& {\it 2.0 } 	&{\it 2.0 }&{\it 2.0 }& 	 &      &{\it 2.0 }&{\it 2.0 }&{\it 2.0 }&{\it 2.0 }\\
{\bf Luminosity}           & {\it 3.8 } & {\it 3.8 } 	& {\it 3.8 } 	&{\it 3.8 }&{\it 3.8 }& 	 &      &{\it 3.8 }&{\it 3.8 }&{\it 3.8 }&{\it 3.8 }\\
{\bf Luminosity Monitor}   & {\it 4.4 } & {\it 4.4 } 	& {\it 4.4 } 	&{\it 4.4 }&{\it 4.4 }& 	 &      &{\it 4.4 }&{\it 4.4 }&{\it 4.4 }&{\it 4.4 }\\
\end{tabular}
\end{ruledtabular}

\end{center}
\end{table}

\begin{table}
\begin{center}
\vskip 0.3cm
{\centerline{CDF: $H\rightarrow W^+W^-\rightarrow\ell^{\pm}\ell^{\prime \mp}$ with two or more associated jets channel relative uncertainties (\%)}}
\vskip 0.0999cm
\begin{ruledtabular}
\begin{tabular}{lccccccccccc} \\
Contribution               &  $WW$    &  $WZ$         &  $ZZ$   &  $t\bar{t}$  &  DY   &  $W\gamma$    & $W$+jet &$gg\to H$&  $WH$ &  $ZH$  &  VBF    \\ \hline
{\bf Cross Section :}      &        	&        	&        	&        &        	&        &         &        &        &        &        \\
Scale                      &        	&        	&        	&        &        	&        &         & 33.0  &        &        &        	\\
PDF Model                  &        	&        	&        	&        &        	&        &         & 29.7  &        &        &        	\\
Total                      & {\it 6.0 }& {\it 6.0 }   & {\it 6.0 }   &  7.0  &  	 	& 	 & 	   &        &{\bf 5.0 }&{\bf 5.0 }& 10.0  \\
{\bf Acceptance :}         &        	&        	&        	&        &        	&        &         &        &        &        &        \\
Scale (jets)               & {\it -8.2 }&        	&        	&        &        	&        &         &       &        &        &        	\\
PDF Model (leptons)        & 		 &  		&  		&	 &	        &  	 &      &{\it 4.8 }&	     &        &         \\
PDF Model (jets)           & {\it 4.2 }&        	&        	&        &        	&        &         & -12.3  &        &        &        	\\
Higher-order Diagrams      &  		& {\it 10.0 }  & {\it 10.0 }  & 10.0  &              & 10.0  & 	   &        &{\bf 10.0 }&{\bf 10.0 }&{\bf 10.0 }\\
$\MET$ Modeling            &  		&   	        &     	        &        & 25.5  	& 	 &         &        &        &        &     	\\
Conversion Modeling      &            &               &               &        &              & 10.0  &         &        &        &        &     	\\
Jet Fake Rates             &        	&        	&        	&        &        	&        & 28.0   &        &        &        &        	\\
Jet Energy Scale& {\it -14.8 } & {\it -12.9 } & {\it -12.1 }  & {\it -1.7 }  & {\it -29.2 } & {\it -22.0 } &  & {\it -17.0 } & {\it -4.0 } & {\it -2.3 } & {\it -4.0 } \\
$b$-tag Veto               &        	&        	&        	&  3.8  &        	&        &         &        &        &        &        	\\
Lepton ID Efficiencies     & {\it 3.8 } &  {\it 3.8 } &  {\it 3.8 }  &{\it 3.8 }&{\it 3.8 }&   	 &      &{\it 3.8 }&{\it 3.8 }&{\it 3.8 }&{\it 3.8 }\\
Trigger Efficiencies       & {\it 2.0 } &  {\it 2.0 } &  {\it 2.0 }  &{\it 2.0 }&{\it 2.0 }&  	 &      &{\it 2.0 }&{\it 2.0 }&{\it 2.0 }&{\it 2.0 }\\
{\bf Luminosity}           & {\it 3.8 } &  {\it 3.8 } &  {\it 3.8 }  &{\it 3.8 }&{\it 3.8 }&  	 &      &{\it 3.8 }&{\it 3.8 }&{\it 3.8 }&{\it 3.8 }\\
{\bf Luminosity Monitor}   & {\it 4.4 } &  {\it 4.4 } &  {\it 4.4 }  &{\it 4.4 }&{\it 4.4 }&  	 &      &{\it 4.4 }&{\it 4.4 }&{\it 4.4 }&{\it 4.4 }\\
\end{tabular}
\end{ruledtabular}
\end{center}
\end{table}

%
%%%% Low Mll
%

\begin{table}
\begin{center}
\caption{\label{tab:cdfsystww4} Systematic uncertainties on the signal and background contributions for CDF's low-$M_{\ell\ell}$
$H\rightarrow W^+W^-\rightarrow\ell^{\pm}\ell^{\prime \mp}$ channel with zero or one associated jets.  This channel is sensitive
to only gluon fusion production.  Systematic uncertainties are listed by name (see the original references for a detailed
explanation of their meaning and on how they are derived).  Systematic uncertainties for $H$ shown in this table are obtained
for $m_H=160$ GeV/$c^2$.  Uncertainties are relative, in percent, and are symmetric unless otherwise indicated.  The uncertainties
associated with the different background and signal processed are correlated within individual categories unless otherwise noted.
In these special cases, the correlated uncertainties are shown in either italics or bold face text.}
\vskip 0.1cm
{\centerline{CDF: low $M_{\ell\ell}$ $H\rightarrow W^+W^-\rightarrow\ell^{\pm}\ell^{\prime \mp}$ with zero or one associated jets channel relative uncertainties (\%)}}
\vskip 0.099cm
\begin{ruledtabular}
\begin{tabular}{lcccccccc} \\
Contribution            & $WW$       & $WZ$       & $ZZ$       & $t\bar{t}$ & DY      & $W\gamma$  & $W$+jet(s) & $gg\to H$ \\ \hline
{\bf Cross Section :}   &            &            &            &            &         &            &            &           \\
Scale                   &            &            &            &            &         &            &            & 12.0       \\
PDF Model               &            &            &            &            &         &            &            & 10.5        \\
Total                   & {\it 6.0 }  & {\it 6.0 }  & {\it 6.0 }  &  7.0 &  5.0  &            &            &            \\
{\bf Acceptance :}      &            &            &            &            &         &            &            &           \\
Scale (jets)            & {\it -0.4} &            &            &            &         &            &            &           \\
PDF Model (leptons)     & 	     &            &            &            &         &            &         & {\it 1.0 }\\
PDF Model (jets)        &{\it 1.6 } &            &            &            &         &            &            & 2.1        \\
Higher-order Diagrams   & 	     &  10.0     &   10.0    &   10.0    & 10.0   &            &            &           \\
Jet Energy Scale           & {\it 1.0 }  & {\it 2.3 } & {\it 2.0 } & {\it 12.9 } & {\it 6.4 } & {\it 1.3 } &            & {\it 2.4 } \\
Conversion Modeling      &            &            &            &            &         & 10.0        &            &           \\
Jet Fake Rates          &            &            &            &            &         &            & 18.4        &           \\
Lepton ID Efficiencies     & {\it 3.8 } &  {\it 3.8 } &  {\it 3.8 } & {\it 3.8 } & {\it 3.8 } &  		  &         & {\it 3.8 } \\
Trigger Efficiencies       & {\it 2.0 } &  {\it 2.0 } &  {\it 2.0 } & {\it 2.0 } & {\it 2.0 } &  		  &         & {\it 2.0 } \\
{\bf Luminosity}           & {\it 3.8 } &  {\it 3.8 } &  {\it 3.8 } & {\it 3.8 } &   {\it 3.8 }    &          &        & {\it 3.8 }  \\
{\bf Luminosity Monitor}   & {\it 4.4 } &  {\it 4.4 } &  {\it 4.4 } & {\it 4.4 } &   {\it 4.4 }    &          &        & {\it 4.4 }  \\
\end{tabular}
\end{ruledtabular}
\end{center}
\end{table}

%%%%%%%%%%%%%%%   CDF Hadronic tau WW channels

\begin{table}
\begin{center}
\caption{\label{tab:cdfsystww5}   Systematic uncertainties on the signal and background contributions for CDF's
$H\rightarrow W^+W^-\rightarrow e^{\pm} \tau^{\mp}$ and $H\rightarrow W^+W^-\rightarrow \mu^{\pm} \tau^{\mp}$
channels.  These channels are sensitive to gluon fusion production, $WH, ZH$ and VBF production.  Systematic
uncertainties are listed by name (see the original references for a detailed explanation of their meaning
and on how they are derived).  Systematic uncertainties for $H$ shown in this table are obtained for
$m_H=160$ GeV/$c^2$.  Uncertainties are relative, in percent, and are symmetric unless otherwise indicated.
The uncertainties associated with the different background and signal processed are correlated within individual
categories unless otherwise noted.  In these special cases, the correlated uncertainties are shown in either
italics or bold face text.}
\vskip 0.1cm
{\centerline{CDF: $H\rightarrow W^+W^-\rightarrow e^{\pm} \tau^{\mp}$ channel relative uncertainties ( )}}
\vskip 0.099cm
\begin{ruledtabular}
\begin{tabular}{lccccccccccccccc} \\
Contribution                 & $WW$  & $WZ$ & $ZZ$ & $t\bar{t}$  & $Z\rightarrow\tau\tau$  & $Z\rightarrow\ell\ell$  & $W$+jet  & $W\gamma$  & $gg\to H$  & $WH$ & $ZH$ & VBF  \\ \hline
Cross section                & 6.0   & 6.0  & 6.0  & 10.0 &  5.0 &  5.0 &       &      &  10.3  &   5   &   5   &  10  \\
Measured W cross-section     &       &      &      &      &      &      & 12    &      &        &       &       &      \\
PDF Model                    & 1.6   & 2.3  & 3.2  & 2.3  &  2.7 &  4.6 & 2.2   & 3.1  &   2.5  & 2.0   &  1.9  & 1.8  \\
Higher order diagrams        & 10    & 10   & 10   & 10   &  10  &  10  &       & 10   &        & 10    &  10   & 10   \\
Conversion modeling          &       &      &      &      &      &      &       & 10   &        &       &       &      \\
Trigger Efficiency           & 0.5   & 0.6  & 0.6  & 0.6  & 0.7  & 0.5  & 0.6   & 0.6  &   0.5  & 0.5   &  0.6  & 0.5  \\
Lepton ID Efficiency         & 0.4   & 0.5  & 0.5  & 0.4  & 0.4  &  0.4 & 0.5   & 0.4  &   0.4  & 0.4   &  0.4  & 0.4  \\
$\tau$ ID Efficiency         & 1.0   & 1.3  & 1.9  & 1.3  & 2.1  &      &       & 0.3  &   2.8  &  1.6  &  1.7  & 2.8  \\
Jet into $\tau$ Fake rate    & 5.8   & 4.8  & 2.0  & 5.1  &      &  0.1 & 8.8   &      &        &  4.2  &  4.0  & 0.4  \\
Lepton into $\tau$ Fake rate & 0.2   & 0.1  & 0.6  & 0.2  &      &  2.3 &       & 2.1  &  0.15  & 0.06  &  0.15 & 0.11 \\
W+jet scale                  &       &      &      &      &      &      & 1.6   &      &        &       &       &      \\
MC Run dependence            & 2.6   & 2.6  & 2.6  &      &      &      & 2.6   &      &        &       &       &      \\
Luminosity                   & 3.8   & 3.8  & 3.8  & 3.8  & 3.8  & 3.8  & 3.8   & 3.8  &  3.8   & 3.8   &  3.8  & 3.8  \\
Luminosity Monitor           & 4.4   & 4.4  & 4.4  & 4.4  & 4.4  & 4.4  & 4.4   & 4.4  &  4.4   & 4.4   &  4.4  & 4.4  \\
%\bf Total                    & 14.7  & 14.4	& 14.0 & 16.4 &	13.1 & 13.7 & 16.5  & 15.8 & 12.5   & 13.6  & 13.5  & 15.7 \\
\end{tabular}
\end{ruledtabular}

\vskip 0.3cm
{\centerline{CDF: $H\rightarrow W^+W^-\rightarrow \mu^{\pm} \tau^{\mp}$ channel relative uncertainties (\%)}}
\vskip 0.099cm
\begin{ruledtabular}
\begin{tabular}{lcccccccccccccc} \\
    Contribution                 & $WW$  & $WZ$ & $ZZ$ & $t\bar{t}$  & $Z\rightarrow\tau\tau$  & $Z\rightarrow\ell\ell$  & $W$+jet  & $W\gamma$  & $gg\to H$  & $WH$ & $ZH$ & VBF  \\ \hline
    Cross section                & 6.0  & 6.0  & 6.0  & 10.0 & 5.0 &  5.0 &     &      & 10.4  &   5   &   5   &  10  \\
    Measured W cross-section     &      &      &      &      &     &      & 12  &      &       &       &       &      \\
    PDF Model                    & 1.5  & 2.1  & 2.9  & 2.1  & 2.5 & 4.3  & 2.0 & 2.9  &  2.6  & 2.2   &  2.0  & 2.2  \\
    Higher order diagrams        & 10   & 10   & 10   & 10   &     &      &     & 11   &       & 10    &  10   & 10   \\
    Trigger Efficiency           & 1.3  & 0.7  & 0.7  & 1.1  & 0.9 & 1.3  & 1.0 & 1.0  &  1.3  & 1.3   &  1.2  & 1.3  \\
    Lepton ID Efficiency         & 1.1  & 1.4  & 1.4  & 1.1  & 1.2 & 1.1  & 1.4 & 1.3  &  1.0  & 1.0   &  1.0  & 1.0  \\
    $\tau$ ID Efficiency         & 1.0  & 1.2  & 1.4  & 1.6  & 1.9 &      &     &      &  2.9  &  1.6  &  1.7  & 2.8  \\
    Jet into $\tau$ Fake rate    & 5.8  & 5.0  & 4.4  & 4.4  &     & 0.2  & 8.8 &      &       &  4.5  &  4.2  & 0.4  \\
    Lepton into $\tau$ Fake rate & 0.06 & 0.05 & 0.09 & 0.04 &     & 1.9  &     & 1.2  & 0.04  & 0.02  &  0.02 & 0.04 \\
    W+jet scale                  &      &      &      &      &     &      & 1.4 &      &       &       &       &      \\
    MC Run dependence            & 3.0  & 3.0  & 3.0  &      &     &      & 3.0 &      &       &       &       &      \\
    Luminosity                   & 3.8  & 3.8  & 3.8  & 3.8  & 3.8 & 3.8  & 3.8 & 3.8  & 3.8   & 3.8   &  3.8  & 3.8  \\
    Luminosity Monitor           & 4.4  & 4.4  & 4.4  & 4.4  & 4.4 & 4.4  & 4.4 & 4.4  & 4.4   & 4.4   &  4.4  & 4.4  \\
%    \bf Total                    & 14.8 & 14.6 & 14.6 & 16.2 & 8.5 & 9.2  & 16.6& 13.0 & 13.6  & 14.0  &  14.0 & 16.5 \\
\end{tabular}
\end{ruledtabular}
\end{center}
\end{table}

%%%%%%%%%%%%%%%   CDF WWW

\begin{table}
\begin{center}
\caption{\label{tab:cdfsystwww} Systematic uncertainties on the signal and background contributions for
CDF's $WH\rightarrow WWW \rightarrow\ell^{\pm}\ell^{\prime \pm}$ channel with one or more associated
jets and $WH\rightarrow WWW \rightarrow \ell^{\pm}\ell^{\prime \pm} \ell^{\prime \prime \mp}$ channel.
These channels are sensitive to only $WH$ and $ZH$ production.  Systematic uncertainties are listed
by name (see the original references for a detailed explanation of their meaning and on how they are
derived).  Systematic uncertainties for $H$ shown in this table are obtained for $m_H=160$ GeV/$c^2$.
Uncertainties are relative, in percent, and are symmetric unless otherwise indicated.  The uncertainties
associated with the different background and signal processed are correlated within individual categories
unless otherwise noted.  In these special cases, the correlated uncertainties are shown in either italics
or bold face text.}
\vskip 0.1cm
{\centerline{CDF: $WH \rightarrow WWW \rightarrow\ell^{\pm}\ell^{\prime\pm}$ channel relative uncertainties (\%)}}
\vskip 0.099cm
\begin{ruledtabular}
\begin{tabular}{lccccccccc} \\
Contribution               & $WW$          &   $WZ$       &  $ZZ$        & $t\bar{t}$   &  DY          & $W\gamma$    & $W$+jet &  $WH$        &  $ZH$        \\ \hline
{\bf Cross Section}        &   {\it 6.0 } &  {\it 6.0 } &  {\it 6.0 } &       7.0  &        5.0  &              &     	&  {\bf 5.0 } &  {\bf 5.0 } \\
Scale (Acceptance)         &  {\it -6.1 } &              &              &              &              &              &         &              &              \\
PDF Model (Acceptance)     &   {\it 5.7 } &  		  &  		 &  		&  	       &  	      &         &  	       & 	      \\
Higher-order Diagrams      &   		   & {\it 10.0 } & {\it 10.0 } &       10.0  &       10.0  &       10.0  &         & {\bf 10.0 } & {\bf 10.0 } \\
Conversion Modeling        &               &              &              &              &              &       10.0  &         &              &              \\
Jet Fake Rates             &               &              &              &              &              &              &  38.5  &              &              \\
Jet Energy Scale           & {\it -14.0 } & {\it -3.9 } & {\it -2.8 } & {\it -0.6 } & {\it -7.7 } & {\it -7.6 } &         & {\it -1.0 } & {\it -0.7 } \\
Charge Mismeasurement Rate &  {\it 19.0 } &              &              & {\it 19.0 } & {\it 19.0 } &              &         &              &              \\
Lepton ID Efficiencies     &   {\it 3.8 } &  {\it 3.8 } &  {\it 3.8 } & {\it 3.8 }  &  {\it 3.8 } &  	      &         &  {\it 3.8 } &  {\it 3.8 } \\
Trigger Efficiencies       &   {\it 2.0 } &  {\it 2.0 } &  {\it 2.0 } & {\it 2.0 }  &  {\it 2.0 } &  	      &         &  {\it 2.0 } &  {\it 2.0 } \\
{\bf Luminosity}           &   {\it 3.8 } &  {\it 3.8 } &  {\it 3.8 } & {\it 3.8 }  &  {\it 3.8 } &   	      &         &  {\it 3.8 } &  {\it 3.8 } \\
{\bf Luminosity Monitor}   &   {\it 4.4 } &  {\it 4.4 } &  {\it 4.4 } & {\it 4.4 }  &  {\it 4.4 } &   	      &         &  {\it 4.4 } &  {\it 4.4 } \\
\end{tabular}
\end{ruledtabular}

\vskip 0.3cm
{\centerline{CDF: $WH\rightarrow WWW \rightarrow \ell^{\pm}\ell^{\prime \pm} \ell^{\prime \prime \mp}$ channel relative uncertainties (\%)}}
\vskip 0.0999cm
\begin{ruledtabular}
\begin{tabular}{lccccccc} \\
Contribution                & $WZ$        & $ZZ$        & $Z\gamma$     & $t\bar{t}$  & Fakes       & $WH$         & $ZH$           \\ \hline
{\bf Cross Section}         & {\it 6.0 } &  {\it 6.0 } &    10.0     &       7.0  &             & {\bf 5.0 }  &   {\bf 5.0 }  \\
Higher-order Diagrams       & {\it 10.0 }& {\it 10.0 } &    15.0     &      10.0  &             & {\bf 10.0 } & {\bf 10.0 }    \\
Jet Energy Scale            &             &              &  {\it -2.7 }&             &             &              &                  \\
Jet Fake Rates              &             &              &              &             & 25.6       &              &                 \\
$b$-Jet Fake Rates          &             &              &              &  27.3      &             &              &                 \\
MC Run Dependence           &             &              &   5.0       &	      &             &  		   &                 \\
Lepton ID Efficiencies      & {\it 5.0 } &  {\it 5.0 } &              & {\it 5.0 } &             & {\it 5.0 }  &   {\it 5.0 }    \\
Trigger Efficiencies        & {\it 2.0 } &  {\it 2.0 } &              & {\it 2.0 } &             & {\it 2.0 }  &   {\it 2.0 }    \\
%Total                      &             &              &              &             &             &              &                  \\
{\bf Luminosity}            & {\it 3.8 } &  {\it 3.8 } &              & {\it 3.8 } &             &  	 {\it 3.8 }&   {\it 3.8 }    \\
{\bf Luminosity Monitor}    & {\it 4.4 } &  {\it 4.4 } &              & {\it 4.4 } &             &  	 {\it 4.4 }&   {\it 4.4 }    \\
\end{tabular}
\end{ruledtabular}

\end{center}
\end{table}

\begin{table}
\begin{center}
\caption{\label{tab:cdfsystzww} Systematic uncertainties on the signal and background contributions for
CDF's $ZH\rightarrow ZWW \rightarrow \ell^{\pm}\ell^{\mp} \ell^{\prime \pm}$ channels with 1 jet and 2
or more jets.  These channels are sensitive to only $WH$ and $ZH$ production.  Systematic uncertainties
are listed by name (see the original references for a detailed explanation of their meaning and on how
they are derived).  Systematic uncertainties for $H$ shown in this table are obtained for $m_H=160$
GeV/$c^2$.  Uncertainties are relative, in percent, and are symmetric unless otherwise indicated.  The
uncertainties associated with the different background and signal processed are correlated within
individual categories unless otherwise noted.  In these special cases, the correlated uncertainties are
shown in either italics or bold face text.}
\vskip 0.1cm
{\centerline{CDF: $ZH\rightarrow ZWW \rightarrow \ell^{\pm}\ell^{\mp} \ell^{\prime \pm}$ with one associated jet channel relative uncertainties (\%)}}
\vskip 0.0999cm
\begin{ruledtabular}
\begin{tabular}{lccccccc} \\
Contribution                & $WZ$        & $ZZ$         & $Z\gamma$    & $t\bar{t}$  & Fakes       & $WH$         & $ZH$           \\ \hline
{\bf Cross Section}         & {\it 6.0 } &  {\it 6.0 } &   10.0      &       7.0  &             & {\bf 5.0 }  &   {\bf 5.0 }  \\
Higher-order Diagrams       & {\it 10.0 }& {\it 10.0 } &    15.0     &      10.0  &             & {\bf 10.0 } & {\bf 10.0 }    \\
Jet Energy Scale            & {\it -7.6 }& {\it -2.3 } &  {\it -5.3 }& {\it  9.4 }&             &  {\it -9.0 }& {\it  8.1 }    \\
Jet Fake Rates              &             &              &              &             & 24.8       &              &                 \\
$b$-Jet Fake Rates          &             &              &              &  42.0      &             &              &                 \\
MC Run Dependence           &             &              &   5.0       &	      &             &  		   &                 \\
Lepton ID Efficiencies      & {\it 5.0 } &  {\it 5.0 } &              & {\it 5.0 } &             & {\it 5.0 }  &   {\it 5.0 }   \\
Trigger Efficiencies        & {\it 2.0 } &  {\it 2.0 } &              & {\it 2.0 } &             &   {\it 2.0 }&   {\it 2.0 }   \\
%Total                      &             &              &              &             &             &              &                  \\
{\bf Luminosity}            & {\it 3.8 } &  {\it 3.8 } &              & {\it 3.8 } &             &  	 {\it 3.8 }&   {\it 3.8 }    \\
{\bf Luminosity Monitor}    & {\it 4.4 } &  {\it 4.4 } &              & {\it 4.4 } &             &  	 {\it 4.4 }&   {\it 4.4\%}    \\
\end{tabular}
\end{ruledtabular}

\vskip 0.3cm
{\centerline{CDF: $ZH\rightarrow ZWW \rightarrow \ell^{\pm}\ell^{\mp} \ell^{\prime \pm}$ with two or more associated jets channel relative uncertainties (\%)}}
\vskip 0.0999cm
\begin{ruledtabular}
\begin{tabular}{lccccccc} \\
Contribution                & $WZ$        & $ZZ$         & $Z\gamma$    & $t\bar{t}$  & Fakes       & $WH$         & $ZH$           \\ \hline
{\bf Cross Section}         & {\it 6.0 } &  {\it 6.0 } &   10.0      &       7.0  &             & {\bf 5.0 }  &   {\bf 5.0 }  \\
Higher-order Diagrams       & {\it 10.0 }& {\it 10.0 } &    15.0     &      10.0  &             & {\bf 10.0 } & {\bf 10.0 }    \\
Jet Energy Scale            &{\it -17.8 }& {\it -13.1 }& {\it -18.2 }& {\it -3.6 }&             & {\it -15.4 }& {\it -4.9 }    \\
Jet Fake Rates              &             &              &              &             & 25.6       &              &                 \\
$b$-Jet Fake Rates          &             &              &              &  22.2      &             &              &                 \\
MC Run Dependence           &             &              &   5.0       &	      &             &  		   &                 \\
Lepton ID Efficiencies      & {\it 5.0 } &  {\it 5.0 } &              & {\it 5.0 } &             & {\it 5.0 }  &   {\it 5.0 }   \\
Trigger Efficiencies        & {\it 2.0 } &  {\it 2.0 } &              & {\it 2.0 } &             &   {\it 2.0 }&   {\it 2.0 }   \\
%Total                      &             &              &              &             &             &              &                  \\
{\bf Luminosity}            & {\it 3.8 } &  {\it 3.8 } &              & {\it 3.8 } &             &  	 {\it 3.8 }&   {\it 3.8 }    \\
{\bf Luminosity Monitor}    & {\it 4.4 } &  {\it 4.4 } &              & {\it 4.4 } &             &  	 {\it 4.4 }&   {\it 4.4 }    \\
\end{tabular}
\end{ruledtabular}

\end{center}
\end{table}

%%%%%%%%%%%%%%%   D0 WW ee, emu and mumu

\begin{table}
\begin{center}
\caption{\label{tab:d0systww} Systematic uncertainties on the signal and background contributions for D0's
$H\rightarrow WW \rightarrow\ell^{\pm}\ell^{\prime \mp}$ channels.  Systematic uncertainties are listed by
name; see the original references for a detailed explanation of their meaning and on how they are derived.
Shape uncertainties are labeled with the ``s'' designation. Systematic uncertainties shown in this table are obtained for the $m_H=165$ GeV/c$^2$ Higgs selection.
Uncertainties are relative, in percent, and are symmetric unless otherwise indicated.}
\vskip 0.1cm
%{\centerline{D0: $H\rightarrow WW \rightarrow e^{\pm} e^{ \mp}$ channel relative uncertainties (\%)}}
{\centerline{D0: $H\rightarrow WW \rightarrow\ell^{\pm}\ell^{\prime \mp}$ channels relative uncertainties (\%)}}
\vskip 0.099cm
\begin{ruledtabular}
\begin{tabular}{ l  c  c  c  c  c  c  c}  \\
Contribution & Diboson & ~~$Z/\gamma^* \rightarrow \ell\ell$~~&$~~W+jet/\gamma$~~ &~~~~$t\bar{t}~~~~$    & ~~Multijet~~  & ~~~~$H$~~~~      \\
\hline
Luminosity/Normalization    &  6      &   6         & 6             & 6            & 30   &   6      \\
Cross Section      &  7        &   5           & 6            & 10           & --  &   7-33         \\
PDF              &2.5           &   2.5          & 2.5             & 2.5     & --   &   8-30      \\
EM Identification       &  2.5      &   2.5          & 2.5            & 2.5      & --   &   2.5        \\
Muon Identification     &  4           &  4           & 4             & 4            & --   &   4       \\
Vertex Confirmation (s) &  2-6      &  1-7       & 1-6      & 1-8            & --   &   1-8       \\
Jet identification (s) &  2-5       &   2-5       & 2-5         & 2-5        & --   &  2-5      \\
Jet Energy Scale (s)   &  2-3         &   1-4        & 1-8          & 1-4         & --   &   1-10      \\
Jet Energy Resolution(s)   &  1-4     &   1-4        & 1-12          & 1-3         & --   &   1-12      \\
B-tagging    &  10           &   10        & 10          & 5          & --   &   10      \\
\end{tabular}
\end{ruledtabular}

\end{center}
\end{table}

%%%%%% D0 mu tau_had

\begin{table}
\begin{center}
\caption{\label{tab:d0systwwtau} Systematic uncertainties on the signal and background contributions for D0's
$H\rightarrow W^+ W^- \rightarrow \mu\nu \tau_{had}\nu $ channel.  Systematic uncertainties are listed by
name; see the original references for a detailed explanation of their meaning and on how they are derived.
Shape uncertainties are labeled with the shape designation (s). Systematic uncertainties shown in this table are obtained for the $m_H=165$ GeV/c$^2$ Higgs selection.
Uncertainties are relative, in percent, and are symmetric unless otherwise indicated.}
\vskip 0.1cm
{\centerline{D0: $H\rightarrow W^+ W^- \rightarrow \mu\nu \tau_{had}\nu $ channel relative uncertainties (\%)}}
\vskip 0.099cm
\begin{ruledtabular}
\begin{tabular}{ l  c  c  c  c  c  c  c}  \\
Contribution       & Diboson    & ~~$Z/\gamma^* \rightarrow
\ell\ell$~~ &$~~W+\tm{jets}$~~ &~~~~$t\bar{t}~~~~$    & ~~Multijet~~
& ~~~~$H$~~~~ \\
\hline
Luminosity ($\sigma_{\tm{inel}}(p\bar{p})$) &4.6   & 4.6   & -  & 4.6    &    -   &   4.6    \\
Luminosity Monitor    &  4.1  & 4.1           & -          & 4.1         &-        &   4.1    \\
Trigger     &  5.0            &5.0             & -          & 5.0        &-        &   5.0    \\
Lepton ID    &  3.7   &3.7           & -           & 3.7              &-        &   3.7    \\
EM veto        &  5.0         &-         & -         & 5.0        &-        &   5.0    \\
Tau Energy Scale (s)    &  1.0        &1.1          & -        & $<$1     &-        &   $<$1   \\
Jet Energy Scale (s)    &  8.0     &  $<$1    & -       & 1.8     &-        &   2.5    \\
Jet identification (s)  &  $<$1       & $<$1       & -         & 7.5        &-        &   5.0    \\
Multijet  (s)  ~~~~~    &  --      & -    & -     & -       &20-50    &   -      \\
Cross Section     &  7.0       & 4.0      & -      & 10       & -        &   10     \\
Modeling    ~~~~~       &  1.0     &-      & 10    & -      &-        &   3.0    \\
\end{tabular}
\end{ruledtabular}
\end{center}
\end{table}

%%%%%%

%%%%%%%%%%%%%%%   D0 WWW

\begin{table}
\begin{center}
\caption{\label{tab:d0systwww} Systematic uncertainties on the signal and background contributions for D0's
$WH \rightarrow WWW \rightarrow\ell^{\prime \pm}\ell^{\prime \pm}$ channel.  Systematic uncertainties are
listed by name; see the original references for a detailed explanation of their meaning and on how they are
derived. Shape uncertainties are labeled with the ``shape'' designation.  Systematic uncertainties for signal
shown in this table are obtained for $m_H=165$ GeV/$c^2$.  Uncertainties are relative, in percent, and are
symmetric unless otherwise indicated.}
\vskip 0.1cm
{\centerline{D0: $VH \rightarrow\ell^{\pm}\ell^{\prime\pm} + X $ Run~IIa channel relative uncertainties (\%)}}
\vskip 0.099cm
\begin{ruledtabular}
\begin{tabular}{ l  c  c  c  c c } \\
Contribution		&WZ/ZZ		&W+jet		&ChargeFlip	&Multijet 		& $VH \rightarrow \ell\ell X$	\\
\hline
Cross section		& 7		& 6		& 0		& 0			& 0		\\
Normalization		& 4		& 4		& 0		& 0			& 0		\\
Trigger ($\mu\mu$)		& 0		& 0		& 0		& 0			& 2		\\
LeptonID ($ee$)		& 8.6		& 8.6 		& 0		& 0			& 8.6		\\
LeptonID ($\mu\mu$)	 	& 4		& 4		& 0		& 0			& 4		\\
LeptonID ($e\mu$)		& 6.3		& 6.3		& 0		& 0			& 6.3		\\
JetID/JES		& 2		& 2		& 0		& 0			& 2	\\
Jet-Lepton Fake		& 0		& 20		& 0		& 0			& 0		\\
Instrumental ($ee$) 	& 0		& 0		& 0		& 52			& 44	\\
Instrumental ($e\mu$ 	& 0		& 0		& 0		& 0			& 29		\\
Instrumental ($\mu\mu$) & 0		& 0		& 0		& 155			& 42		\\
Instrumental Model	& -		& -		& shape		& shape			& -		\\
\end{tabular}
\end{ruledtabular}

\vskip 0.3cm
{\centerline{D0: $VH \rightarrow\ell^{\pm}\ell^{\prime\pm} + X $ Run~IIb channel relative uncertainties (\%)}}
\vskip 0.099cm
\begin{ruledtabular}
\begin{tabular}{ l  c  c  c  c c } \\
Contribution 	&WZ/ZZ		&W+jet		&ChargeFlip	&Multijet 		& $VH \rightarrow \ell\ell X$	\\
\hline
Cross section		& 7		& 6		& 0			& 0			& 0		\\
Normalization		& 4		& 4		& 0			& 0			& 0		\\
Trigger ($\mu\mu$)		& 0		& 0		& 0			& 0			& 5		\\
LeptonID ($ee$)		& 8.6		& 8.6 		& 0			& 0			& 8.6		\\
LeptonID ($\mu\mu$)	 	& 4		& 4		& 0			& 0			& 4		\\
LeptonID ($e\mu$)		& 6.3		& 6.3		& 0			& 0			& 6.3		\\
JetID/JES		& 2		& 2		& 0			& 0			& 2	\\
Jet-Lepton Fake		& 0		& 20		& 0			& 0			& 0		\\
Instrumental ($ee$)	& 0		& 0		& 0			& 23			& 31		\\
Instrumental ($e\mu$)	& 0		& 0		& 0			& 0			& 19		\\
Instrumental ($\mu\mu$)	& 0		& 0		& 0			& 43			& 28		\\
Instrumental Model	& -		& -		& shape			& shape			& -		\\
\end{tabular}
\end{ruledtabular}
\end{center}
\end{table}

%%%%%%%%%%%%%%%%%%%%%%%%%%%%% HW -> lnujj

\begin{table*}
\begin{center}
\caption{\label{tab:d0lvjj}
%Systematic uncertainties for the electron and muon channels. 
Systematic uncertainties on the signal and background contributions for D0's
 $H\rightarrow W W^{*} \rightarrow \ell\nu jj$ electron and muon channels.  Systematic uncertainties are listed
 by name; see the original references for a detailed explanation of their meaning and on how they are
 derived.
Signal uncertainties are shown for $m_H=160$ GeV/$c^2$ for all channels except for $WH$, 
shown for $m_H=115$ GeV/$c^2$.  Those affecting the shape of 
the RF discriminant are indicated with ``Y.''
Uncertainties are listed as relative changes in normalization, 
in percent, except for those also marked by ``S,'' where 
the overall normalization is constant, and the value given
denotes the maximum percentage change from nominal in any region of the
distribution.}

\vskip 0.1cm
{\centerline{D0: $H\rightarrow W W^{*} \rightarrow \ell\nu jj$ Run~II channel relative uncertainties (\%)}}
\vskip 0.099cm
\begin{ruledtabular}
\begin{tabular}{llccccccl}

Contribution & Shape & $W$+jets & $Z$+jets & Top & Diboson & $gg\to H$ & $qq\to qqH$ & $WH$ \\ \hline
Jet energy scale & Y & $\binom{+6.7}{-5.4}^S$ & $<0.1$ & $\pm$0.7 & $\pm$3.3 & $\binom{+5.7}{-4.0}$ & $\pm$1.5 &$\binom{+2.7}{-2.3}$  \\ 
Jet identification & Y & $\pm 6.6^S$ & $<0.1$ & $\pm$0.5 & $\pm$3.8  & $\pm$1.0 & $\pm$1.1 & $\pm$1.0 \\ 
Jet resolution & Y & $\binom{+6.6}{-4.1}^S$ & $<0.1$ & $\pm$0.5 & $\binom{+1.0}{-0.5}$ & $\binom{+3.0}{-0.5}$ & $\pm 0.8$ & $\pm 1.0$ \\ 
Association of jets with PV & Y & $\pm 3.2^S$ & $\pm 1.3^S$ & $\pm$1.2 & $\pm$3.2 & $\pm$2.9 & $\pm$2.4 & $\binom{+0.9}{-0.2}$ \\ 
Luminosity & N & n/a & n/a & $\pm$6.1 & $\pm$6.1 & $\pm$6.1 & $\pm$6.1 &  $\pm$6.1 \\ 
Muon trigger  & Y & $\pm 0.4^S$ & $<0.1$ & $<0.1$ & $<0.1$ & $<0.1$ & $<0.1$ &  $<0.1$ \\ 
Electron identification & N & $\pm$4.0  & $\pm$4.0  & $\pm$4.0  & $\pm$4.0  & $\pm$4.0  & $\pm$4.0  & $\pm$4.0 \\ 
Muon identification  & N & $\pm$4.0  & $\pm$4.0  & $\pm$4.0  & $\pm$4.0  & $\pm$4.0  & $\pm$4.0  & $\pm$4.0  \\ 
ALPGEN tuning & Y & $\pm 1.1^S$ & $\pm 0.3^S$ & n/a & n/a & n/a & n/a & n/a \\ 
Cross Section & N & $\pm$6 & $\pm$6 &  $\pm$10 & $\pm$7 & $\pm$10 & $\pm$10 & $\pm$6 \\ 
Heavy-flavor fraction  & Y & $\pm$20 & $\pm$20 & n/a & n/a & n/a & n/a & n/a  \\ 
PDF & Y & $\pm 2.0^S$ & $\pm 0.7^S$ & $<0.1^S$ & $<0.1^S$ & $<0.1^S$ & $<0.1^S$ & $<0.1^S$ \\ 
 &  &  &  &  &  &  &  &  \\ 
 &  & \multicolumn{ 3}{c}{Electron channel} & \multicolumn{ 3}{c}{Muon channel} &  \\ 
Multijet Background & Y  & \multicolumn{ 3}{c}{$\pm$6.5} & \multicolumn{ 3}{c}{$\pm$26} &  \\ 
\end{tabular}
\end{ruledtabular}
\end{center}
\end{table*}

%********* VH ******

\begin{table}
\begin{center}
\caption{\label{tab:d0sysVHtau}  Systematic uncertainties on the signal and background contributions for 
D0's 
%$XH\rightarrow \ell \tau$jj Run~IIb channel ($l=e,\mu$). 
$\tau \tau jj$ Run~IIb channel. 
Systematic uncertainties for the 
Higgs signal shown in this table are obtained for $m_H=135$ GeV/$c^2$.  Systematic uncertainties are listed 
by name; see the original references for a detailed explanation of their meaning and on how they are derived.  
Uncertainties are relative, in percent, and are symmetric unless otherwise indicated. A systematic is denoted 
as flat if it affects the normalization only, and as 'shape' otherwise.}
\vskip 0.1cm
%{\centerline{D0: $XH\rightarrow \mu \tau_{had} jj$ Run~IIb channel relative uncertainties (\%)}}
{\centerline{D0: $\mu \tau_{had} jj$ Run IIb channel relative uncertainties (\%)}}
\vskip 0.099cm
\begin{ruledtabular}
\begin{tabular}{lccccccccc}\\
Contribution &$VH$ Signal &$VBF$ Signal &$GGF$ Signal  &$W+jets$ & $Z+jets$ & Top& diboson& Multijet\\ \hline

Luminosity (D0 specific)  & 4.1 & 4.1 & 4.1 & 4.1 & 4.1 & 4.1 &4.1 &- \\ 
Luminosity (Tevatron common)  & 4.6 & 4.6 & 4.6 & 4.6 & 4.6 & 4.6 &4.6 &- \\ 
$\mu$ ID & 2.9 & 2.9 & 2.9 & 2.9 & 2.9 & 2.9 &2.9 &- \\ 
$\mu$ trigger & 8.6 & 8.6 & 8.6 & 8.6 & 8.6 & 8.6 &8.6 &- \\ 
$\tau$ energy correction & 9.8 & 9.8 & 9.8 & 9.8 & 9.8 & 9.8 &9.8 &- \\ 
$\tau$ track efficiency & 1.4 & 1.4 & 1.4 & 1.4 & 1.4 & 1.4 &1.4 &- \\ 
$\tau$ selection by type & 12,4.2,7 & 12,4.2,7 & 12,4.2,7 & 12,4.2,7 & 12,4.2,7 & 12,4.2,7 &12,4.2,7 &- \\ 
Cross section & 6.2 &4.9& 33 &6.0&6.0&10.0&7.0&-\\
GGF Signal PDF &-&-&29&-&-&-&-&-\\
GGF $H p_T$ Reweighting (Shape)& 1.0 & 1.0 & 1.0 & 1.0 & 1.0 & 1.0 &1.0 &-\\
Vertex confirmation for jets  & 4.0 & 4.0 & 4.0 & 4.0 & 4.0 & 4.0 &4.0 &- \\

Jet ID(Shape)  & $\sim$10 & $\sim$10 & $\sim$10 & $\sim$10& $\sim$10 & $\sim$10 &$\sim$10 &- \\
Jet Energy Resolution (Shape)  & $\sim$10 & $\sim$10 & $\sim$10 & $\sim$10& $\sim$10 & $\sim$10 &$\sim$10 &- \\
Jet energy Scale (Shape)  & $\sim$15 & $\sim$15 & $\sim$15 & $\sim$15& $\sim$15 & $\sim$15 &$\sim$15 &- \\

Jet pT  & 5.5 & 5.5 & 5.5 & 5.5 & 5.5 & 5.5 &5.5 &- \\
PDF reweighting  & 2 & 2 & 2 & 2 & 2 & 2 &2 &- \\
Multijet Normalization  & - & - & - & - & - & - &- &5.3 \\
Multijet Shape   & - & - & - & - & - & - &- &$\sim$15 \\

\end{tabular}
\end{ruledtabular}

\vskip 0.3cm
%{\centerline{D0: $XH\rightarrow e \tau_{had} jj$ Run~IIb relative uncertainties (\%)}}
{\centerline{D0: $e \tau_{had} jj$ Run IIb relative uncertainties (\%)}}
\vskip 0.099cm
\begin{ruledtabular}
\begin{tabular}{lccccccccc}\\
Contribution &$VH$ Signal &$VBF$ Signal &$GGF$ Signal  &$W+jets$ & $Z+jets$ & Top& diboson& Multijet\\ 
\hline
Luminosity (D0 specific)  & 4.1 & 4.1 & 4.1 & 4.1 & 4.1 & 4.1 &4.1 &- \\ 
Luminosity (Tevatron common)  & 4.6 & 4.6 & 4.6 & 4.6 & 4.6 & 4.6 &4.6 &- \\ 
EM ID & 4 & 4 & 4 & 4 & 4 & 4 &4 &- \\ 
e trigger & 2 & 2 & 2 & 2 & 2 & 2 &2 &- \\ 
$\tau$ energy correction & 9.8 & 9.8 & 9.8 & 9.8 & 9.8 & 9.8 &9.8 &- \\ 
$\tau$ track efficiency & 1.4 & 1.4 & 1.4 & 1.4 & 1.4 & 1.4 &1.4 &- \\ 
$\tau$ selection by type & 12,4.2,7 & 12,4.2,7 & 12,4.2,7 & 12,4.2,7 & 12,4.2,7 & 12,4.2,7 &12,4.2,7 &- \\ 
Cross section & 6.2 &4.9& 33 &6.0&6.0&10.0&7.0&-\\
GGF Signal PDF &-&-&29&-&-&-&-&-\\
GGF $H p_T$ Reweighting (Shape)& 1.0 & 1.0 & 1.0 & 1.0 & 1.0 & 1.0 &1.0 &-\\
Vertex confirmation for jets  & 4.0 & 4.0 & 4.0 & 4.0 & 4.0 & 4.0 &4.0 &- \\ 

Jet ID(Shape)  & $\sim$10 & $\sim$10 & $\sim$10 & $\sim$10& $\sim$10 & $\sim$10 &$\sim$10 &- \\
Jet Energy Resolution (Shape)  & $\sim$10 & $\sim$10 & $\sim$10 & $\sim$10& $\sim$10 & $\sim$10 &$\sim$10 &- \\
Jet energy Scale (Shape)  & $\sim$15 & $\sim$15 & $\sim$15 & $\sim$15& $\sim$15 & $\sim$15 &$\sim$15 &- \\

Jet pT  & 5.5 & 5.5 & 5.5 & 5.5 & 5.5 & 5.5 &5.5 &- \\
PDF reweighting  & 2 & 2 & 2 & 2 & 2 & 2 &2 &- \\
Multijet Normalization  & - & - & - & - & - & - &- &4.7 \\
Multijet Shape   & - & - & - & - & - & - &- & $\sim$15 \\

\end{tabular}
\end{ruledtabular}

\end{center}
\end{table}

%******* H -> gg *******

\begin{table}
\begin{center}
\caption{\label{tab:d0systgg} Systematic uncertainties on the signal and background contributions for D0's
$H\rightarrow \gamma \gamma$ channel. Systematic uncertainties for the Higgs signal shown in this table are
obtained for $m_H=125$ GeV/$c^2$.  Systematic uncertainties are listed by name; see the original references
for a detailed explanation of their meaning and on how they are derived.  Uncertainties are relative, in
percent, and are symmetric unless otherwise indicated.}
\vskip 0.1cm
{\centerline{D0: $H \rightarrow \gamma \gamma$ channel relative uncertainties (\%)}}
\vskip 0.099cm
\begin{ruledtabular}
\begin{tabular}{lcc}\\
Contribution &  ~~~Background~~~  & ~~~Signal~~~    \\
\hline
Luminosity~~~~                            &  6     &  6    \\
Acceptance                                &  --    &  2    \\
electron ID efficiency                    &  2     &  --   \\
electron track-match inefficiency         & 10     &  --   \\
Photon ID efficiency                      &  3     &   3   \\
Photon energy scale                       &  2     &   1   \\
%Cross Section ($Z$)                       &  4     &  10   \\
Cross Section                             &  4     &  10   \\
Background subtraction                    &  15 &  -       \\
\end{tabular}
\end{ruledtabular}
\end{center}
\end{table}


\begin{thebibliography}{000}
\bibitem{dzHWWtau} \DZero Collaboration,
`` Search for the Standard Model Higgs boson in $ \mu + \tau_{had} \le 1$ jet final
state  with 7.3 fb$^{-1}$ of data'',
D0 Conference Note 6179.
\end{thebibliography}
\end{document}